\documentclass[usenatbib,useAMS]{mnras}
\usepackage{newtxtext,newtxmath}
\usepackage[utf8]{inputenc}
\usepackage{graphics,epsfig,psfig}
\usepackage{graphicx}
\usepackage[normalem]{ulem}
\usepackage[dvipsnames]{color}
\usepackage{xcolor,subfig}
\usepackage{lipsum}
\usepackage{hyperref}
\usepackage{makecell}

\def \be{\begin{equation}}
\def \ee{\end{equation}}
\def \ba{\begin{eqnarray}}
\def \ea{\end{eqnarray}}

\def \seven{`SF, instantaneous $10^7$ M$_{\odot}$'\,}
\def \eight{`SF, instantaneous $10^8$ M$_{\odot}$'\,}
\def \nine{`SF, instantaneous $10^9$ M$_{\odot}$'\,}
\def \cont{`SF, continuous $2$ M$_{\odot}$ yr$^{-1}$'\,}
\def \agninst{`AGN, instantaneous'\,}
\def \agncont{`AGN, continuous'\,}

\newcommand{\ergps}{erg s$^{-1}$}

\newcommand{\pcc}{ cm$^{-3}$}
\newcommand{\mpcc}{$m_{\rm p}$ cm$^{-3}$}
\newcommand{\kmps}{km s$^{-1}$}
\newcommand{\msun}{M$_\odot$}

\newcommand{\divr}{ {\nabla}\cdot}
\newcommand{\erosita}{\textit{eROSITA} }
\newcommand{\athena}{\textit{Athena} }
\newcommand{\isothermal}{\texttt{isothermal}}
\newcommand{\isentropic}{\texttt{isentropic}}
\newcommand{\coolingflow}{\texttt{rotating cooling-flow}}

\definecolor{capri}{rgb}{0.0, 0.75, 1.0}

\title[CGM outflow interaction]{X-ray signatures of galactic outflows into the circumgalactic medium}
\author[Jana \textit{et. al.}]
{Ranita Jana$^1$\thanks{Email: ranitajana@mail.tau.ac.il}, Kartick C. Sarkar$^{1,2}$, Jonathan Stern$^{1}$, Amiel Sternberg$^{1,3,4}$\\
$^1$ School of Physics and Astronomy, Tel Aviv University, Ramat Aviv 69978, Israel\\
$^2$ Dept of Space, Planetary and Astronomical Sciences and Engineering (SPASE), Indian Institute of Technology Kanpur, India, 208016\\
$^3$ Center for Computational Astrophysics, Flatiron Institute, 162 Fifth Avenue, New York, NY 10010, USA\\
$^4$ Max Planck Institute for Extraterrestrial Physics, Garching, Germany
}

\begin{document}
\maketitle
\label{firstpage}

\begin{abstract}
We present a set of controlled hydrodynamical simulations to study the effects of strong galactic outflows on the density and temperature structures, and associated X-ray signatures, of extra-planar and circumgalactic gas. We consider three initial state models, isothermal, isentropic, and rotating cooling-flow, for the hot circumgalactic medium (CGM) into which the outflows are driven. The energy sources are either stellar winds and supernovae, or active galactic nuclei. We consider energy injection rates in the range $10^{40} < \dot{E}_{\rm inj} <10^{44.5}$ erg s$^{-1}$, and compute the time-dependent soft X-ray ($0.5-2$ keV) surface brightness. For $\dot{E}_{\rm inj} \gtrsim 10^{41} - 10^{42}$ erg s$^{-1}$, with the exact threshold depending on the initial CGM state, the X-ray response is dominated by dense hot gas in the forward shock that eventually fades into the CGM as a sound wave. The shock surrounds an inner hot bubble leading to a radial flattening of the X-ray surface brightness. For lower energy injection rates, the X-ray surface brightness of the initial CGM state is almost unaffected. We present analytic approximations for the outflow shock propagation and the associated X-ray emissions.
\end{abstract}


\section{Introduction}
\label{sec:intro}
The existence of hot ($\sim 10^6$ K) and extended gaseous coronae or the circumgalactic medium (CGM), surrounding individual galaxies was proposed many decades ago \citep{Spitzer1956, Munch1961, White1978}. Around the Milky Way, validation was obtained via head-tail shapes of high-velocity \ion{H}{i} clouds, \citep{Peek2007, Putman2011}, lack of \ion{H}{i} in dwarf spheroidal galaxies \citep{Blitz2000} and heavy element absorption lines in the spectra of background quasars \citep{Bahcall1969, Thom2012, Tumlinson2017}. Evidence for hot gas around external galaxies is extended X-ray emission halos around massive field galaxies \citep{Anderson2011, Dai2012, Anderson2013, Bogdan2013}, as well as via thermal Sunyaev-Zeldovich CMB distortions \citep{Bregman022,Das2023,Oren2024}. Understanding the formation and evolution of the CGM is important because the CGM serves as a bath of gaseous material that regulates star formation in a galaxy, mediates metal enrichment in the universe \citep{White&Frenk1991, Tegmark1993, Nath1997, Somerville1999, Cole2000, Nelson2013, Hafen2020, MLi2020}, and potentially affects the morphology of blue galaxies \citep{Sales2012, Stern2021, Hafen2022, Yu2023}. Gaseous outflows driven by supernovae \citep[SNe;][]{Ferrara2000, Oppenheimer2006, Hopkins2012, Sarkar2015, Suresh2015, Girichidis2016, Fielding2017, LiBryan2017, Kim2017, MLi2020} and supermassive black holes \citep[SMBH;][]{Nelson2019, Pillepich2021, Truong2023} carry significant amounts of thermal and kinetic energy that can heat up the CGM as part of feedback loops that control the growth of galaxies.

Understanding the processes of energy transfer from within galaxies into the CGM is not well understood due to complexities in the interactions between the outflows and the CGM. Furthermore, there remains substantial uncertainty regarding the specific mechanism responsible for initiating the outflow, whether driven by SNe or SMBH. The amount of energy released during these processes, as well as its distribution between thermal and kinetic forms and its burstiness, is yet to be ascertained. A main issue has been to combine small scales ($\sim$ pc) that can resolve SNe and active galactic nuclei (AGN) interactions with local interstellar matter (ISM) and then to large scales ($\sim 100$ kpc) that can account for energy flows into the CGM. Recent efforts \citep{Creasey2013, Sarkar2015, Martizzi2016, Kim2017, Miao2017, Fielding2018, Hu2019, MLi2020, Kelly2021, Steinwandel2022} suggest that most of the energy from SNe is carried out by the hot ($\gtrsim 10^6$ K) gas. Hence, probing the hot gas, mainly via diffuse X-ray emission, is expected to provide us with crucial information about the energy contained in such outflows.

Connecting the X-ray emission close ($\sim 100$ pc) to the galaxy and to the outflows can be confusing since emission from individual SNe or high/low mass X-ray binaries can contribute significantly. Observations of X-ray emission from the central star-forming regions (due to clustered supernovae; \citealt{Chevalier1985, Yadav2017}) could, in principle, provide us important information regarding the energy production rate in the winds but the X-ray emissions from these regions are either blocked by the ISM of the host galaxy (e.g. in M82) or distorted by the X-ray binaries \citep{Grimm2003, Gilfanov2004, Mineo2012}. Moreover, subsequent radiative cooling, due to multi-phase interactions \citep{Cooper2009, Kwak2010, Fielding2020} as the hot gas propagates outward from the galaxy, changes the energy contained in the wind. Extra-planar ($100$ pc $\lesssim z \lesssim 5$ kpc, where z is the height from the galactic disk) X-ray emission from edge-on galaxies, on the other hand, is free from such confusion with the galactic sources \citep{Strickland2004, Li2013, Wang2016}. The X-ray emission from the wind, however, is affected by multi-phase interactions via bow shocks (\citealt{Cooper2008, Melioli2013, Vijayan2018, Schneider2020, Nguyen2021, Fielding2022}), or non-spherical geometry \citep{Nguyen2022} or via non-equilibrium ionization physics \citep{Gray2019, Sarkar2022}, and is currently under investigation. At even larger distances ($5$ kpc $\lesssim z \lesssim 50$ kpc), the interaction between the outflow and the CGM becomes important in determining the X-ray emission as the outflows shock heat the medium and change its thermal properties \citep{Nelson2015, Sarkar2015, Suresh2015, Fielding2017}. This region is advantageous for studying the impact of the outflow on the large-scale CGM since the X-ray emission is mainly driven by macroscopic processes, such as shocks, that are relatively better quantified. The emission, however, is often contaminated by the initial CGM of the galaxy at these scales \citep{Sarkar2016}. 
 
Up until recently X-ray ($0.5 - 2$ keV) CGM emission has been detected only in relatively massive ($M_\star \gtrsim 10^{11}$ \msun) galaxies \citep{Brown2001, Tullmann2006, Anderson2011, Anderson2013, Li2016, Li2017, Das2019}. However, recent observations using the \erosita telescope have unveiled such X-ray CGM emissions in galaxies of lower mass. \cite{Comparat2022} produced a stacked radial X-ray intensity profile of $40\,000$ galaxies with stellar masses in the range of $8.5 \lesssim M_{\star}/M_{\odot} \lesssim 12$. For Milky Way-type galaxies ($M_\star \sim 10^{10.7}$ \msun), the radial X-ray intensity profile is detectable up to $\sim 100$ kpc from the center \citep[see also][]{Zhang2024}. Another recent paper on \erosita eFEDS observations by \citet{Chadayammuri2022} compared the observed X-ray intensity profile around galaxies with the TNG and EAGLE cosmological simulations. They found that the observed X-ray intensity profile around the star-forming galaxies is shallower in the inner $\sim 30$ kpc region compared to the simulations. The difference may arise due to uncertainties in the baryon fractions of the CGM in TNG and EAGLE \citep{Oppenheimer2021, Crain2023,Oren2024}. It is, however, difficult to determine the main reason for such behavior in global simulations due to the complex coupling of multiple parameters in the numerical evolution. In controlled simulations on the other hand we have the advantage of being able to adjust the parameters manually to build up an understanding of the perturbed systems. In this paper, we perform controlled simulations of Milky-Way type galaxies with our aim of understanding how SNe/SMBH-driven outflows from these galaxies impact the CGM and related X-ray emission properties. We also provide approximate analytical formulae to mimic such interaction for any general outflow/CGM combination.

Our paper is organized as follows. In section \ref{sec:setup}, we describe our simulation setup, i.e. the initial and the boundary conditions. Section \ref{sec:results} describes our results regarding the impact of outflows in the CGM and their X-ray signatures. An approximate analytical technique is presented in section \ref{sec:analytical} for general use with any combination of CGM/outflow parameters. In section \ref{sec:discussion}, we discuss possible implications of our results, and finally, provide the main conclusions in section \ref{sec:summary}.

\begin{figure*}
    \includegraphics[width=\textwidth]{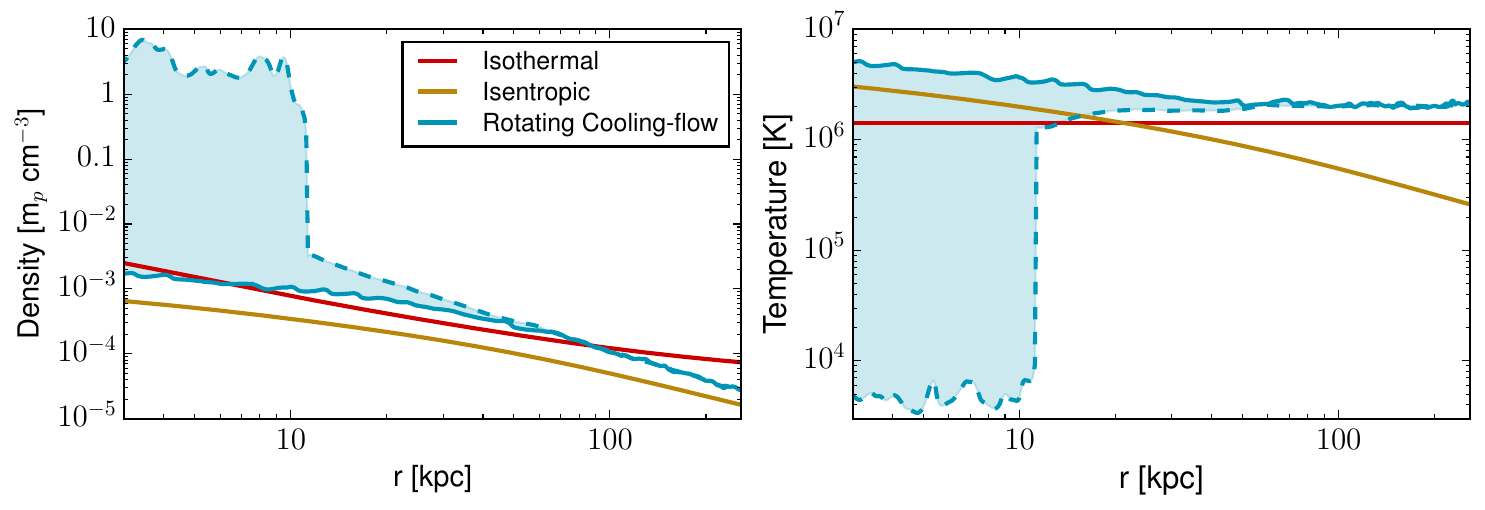}
    \caption{Different initial conditions explored in our simulations. Panels show gas density (left) and temperature (right). The \isothermal\, and \isentropic\, profiles are spherically symmetric. The \coolingflow\, profile depends on the polar angle at small radii, so the density and temperature span a range at each radius with the solid border corresponding to the rotation axis and the dashed border to the galactic plane.
    }
    \label{fig:initial}
\end{figure*}

\section{Simulation set-up}
\label{sec:setup}
We perform a hydrodynamic (HD) simulation in spherical coordinates for a Milky-Way-sized galaxy using the code \textsc{pluto} (\citealt{Mignone2007}). The simulations solve the hydrodynamic equations in $r,\theta$ coordinates assuming axisymmetry, but retain the $\phi$ component of the velocity and associated centrifugal force. We solve the following hydro-dynamical equations corresponding to the conservation of mass, momentum, and energy respectively: 
\begin{eqnarray}
\label{eq:masscont}
&&\frac{\partial \rho}{\partial t} + \nabla\cdot (\rho \boldsymbol{v}) = S_{\rho} \\
&&\frac{\partial}{\partial t} \left(\rho {\boldsymbol v}\right) + \divr ({\rho {\boldsymbol v} \otimes {\boldsymbol v}}+ p {\bf I} )  - \rho \mathbf{g} = 0  \nonumber \\
&&\frac{\partial }{\partial t}\left(\frac{\rho v^2}{2} + e\right) + \divr \left[\left(\frac{\rho v^2}{2}+ e + p\right) {\boldsymbol v}\right]  = - q_{\rm cool}  + S_{\rm e} \nonumber
\end{eqnarray}
Here $\rho$ is the mass density, $\bf{v}$ is the velocity of the gas, $p$ is the pressure, $\rho v^2/2$ is the kinetic energy density, $e = p/(\gamma-1)$ is the thermal energy density and $\gamma = 5/3$ is the adiabatic index. The source terms, $S_{\rm \rho}$ and $S_{\rm e}$ are the injected mass and energy density per unit time. The term, $q_{\rm cool}$, denotes the energy loss rate per unit volume due to radiative cooling and is calculated as discussed in section \ref{subsec:cooling}. 

\subsection{Initial Conditions}
\label{subsec:ini_dist}
The diffuse soft X-ray emission around the galaxy depends on the density, temperature, and metallicity distribution of the CGM. The initial CGM gas distribution also plays a significant role in determining the morphology of the galactic wind and the subsequent CGM-wind interaction. We assume three different initial CGM gas distributions.

\begin{itemize}

\item \underline{\texttt{Rotating Cooling-flow} model:} \cite{Stern2019,Stern2020} demonstrated that in the absence of ongoing heating by feedback, the CGM forms a hot inflow similar to classic `cooling flow' solutions developed for the intracluster medium (ICM), where radiative losses are compensated by compressive heating in the inflow. These cooling flow solutions thus provide a benchmark to test the effects of feedback heating on the CGM as done in this work. Specifically, we use the rotating cooling flow solution in \cite{Stern2023} that extends the classic spherical solutions to axisymmetric solutions which account for angular momentum. Angular momentum support becomes significant near the galaxy and causes the hot inflow to transition into a disk geometry that subsequently cools onto the ISM. The two parameters of this solution are (1) the mass inflow rate and (2) the specific angular momentum of the hot gas as a function of polar angle $\theta$. We set these to $\dot{M}_{\rm if} = 1 \, {\rm M}_{\odot} \rm yr^{-1}$ and $j(\theta)=2000\sin^2\theta$ kpc km s$^{-1}$ as expected for the Milky-Way\footnote{A fully analytic solution for rotating cooling-flow is not yet available, so the initial conditions are derived numerically. The simulation is initialized with an approximate rotating cooling-flow solution and run for $250$ Myr without feedback, during which the hot CGM settles on the correct solution. See \cite{Stern2023} for additional details.}.

\item \underline{\texttt{Isothermal} model:} Combining the \ion{O}{vi} absorption line data of external galaxies observed in the COS-Halos survey \citep{Tumlinson2011} with the Milky Way (MW) X-ray \ion{O}{vii} and \ion{O}{viii} absorption and emission data, \citet{Faerman2017}  provided an analytical hydrostatic model for CGM gas around MW-type galaxies. This model contains a volume-filling hot ($T \approx 1.4 \times 10^6$ K) phase and a warm ($T \sim 10^5$ K) component. The total pressure ($p_{\rm tot}$) has a thermal ($p_{\rm th}$) and a turbulent ($p_{\rm turb}$) component with a ratio defined as $\alpha_{\rm OML}= p_{\rm tot}/p_{\rm th}= 2.1$ independent of radius. Since the volume filling fraction of the warm component is very small ($\sim 0.02$) within the inner 50 kpc of CGM, we assume that the model consists of only a hot phase with a constant temperature, $T = 1.4\times 10^6$ K. We also assume that the gas behaves as a thermal fluid with an adiabatic index $\gamma = 5/3$ but that the total gas pressure is the same as in the original model. The corresponding gas temperature is calculated as $T =  (p_{\rm tot}/\alpha_{\rm OML})/n k_{\rm B}$. 

\item \underline{\texttt{Isentropic} model:} In a follow-up study, \citet{Faerman2020} put forward another hydrostatic CGM model with an alternate adiabatic equation of state. This model is also constrained by the absorption measurements of highly ionized oxygen ions (\ion{O}{vi}-\ion{O}{viii}). The total pressure in this model has three components, i.e. thermal ($p_{\rm th}$), non-thermal ($p_{\rm nth}$), and turbulent ($p_{\rm turb}$) pressure. The entropy parameter ($K = p/\rho^{\gamma}$) for each component remains constant in the initial distribution. The ratio of the total pressure to the thermal pressure, $\alpha_{\rm OML} = p_{\rm tot}/p_{\rm th}$, is equal to 3.2 at $r_{\rm vir}$ and decreases towards inner radii. 

\end{itemize}
We note that the \isothermal\, and \isentropic\, models require a feedback energy injection rate of $\approx10^{41}$ erg s$^{-1}$ on a timescale of $\gtrsim$Gyr, to prevent a cooling flow from developing. For these initial conditions, the short-term feedback (a few hundred Myr) modeled in this paper is assumed to be in addition to the persistent feedback required by these two models.

The density and temperature profiles of the CGM  for each model are shown in Figure \ref{fig:initial}. In the \coolingflow\, model the density and temperature have a dependence on the polar angle at small radii, hence the profiles cover a range at each radius. The solid curve represents the profile along the rotation axis and the dashed curve represents the profile along the galactic plane. We find that the gas distribution is reasonably stable for all the CGM models up to $t \sim 1\, \rm Gyr$, longer than the $\sim 100$ Myr timescale for feedback effects explored in this work.

\subsection{Gravitation force}\label{sec:g_force}
We consider a static gravitational force $g(r)$ relevant for the Milky Way galaxy. For the \isothermal\, and \isentropic\, CGM the gravitational force is calculated following \citet{Klypin2002} (details given in Appendix \ref{sec:emissivity_cloudy}). The gravitational force used for the \coolingflow\, model is similar to a singular isothermal sphere with constant circular velocity ($v_{\rm c} = 200$ km s$^{-1}$) given by $\frac{d\Phi(r)}{dr} = \frac{v_{\rm c}^2}{r}$. The simulation setup neglects the self-gravitational effects inherent to the gaseous medium.

\subsection{Simulation grid}
\label{subsec:grid}
We define our computational box from $0.2$ kpc ($0.4$ kpc for \coolingflow) to $258$ kpc ($r_{\rm vir}$) in the $r$ direction and from $0.01$ to $\pi/2$ in the $\theta$ direction. In the $r$ direction, we define a logarithmic grid with $512$ grid points with $\Delta r \sim 3$ pc at the inner boundary, $0.1$ kpc at $10$ kpc,  $0.4$ kpc at $30$ kpc and $4$ kpc at the outer boundary. For $\theta$, we assume a uniform grid distribution with $256$ points. For $r$, the inner and outer boundaries are set with outflow boundary conditions (i.e. zero gradient condition). For $\theta$, both the inner and outer boundaries are set to be reflective (i.e. only $v_\theta$ changes sign across the boundary). \textsc{pluto} solves the hydrodynamical equations using the finite volume method and solves the Riemann problem in approximation at the cell boundaries to obtain the flux for each variable using the Godunov scheme \citep{Godunov1959}. We employ the \textsc{hllc} Riemann solver \citep{Toro1994} for that purpose and take piecewise linear spatial reconstruction for all variables. The evolution in time is solved using the 2nd-order Runge-Kutta scheme. 

\subsection{Cooling}
\label{subsec:cooling}
We have considered radiative cooling of the CGM gas using the tabulated cooling function calculated from \textsc{cloudy-v22}\footnote{\href{https://gitlab.nublado.org/cloudy/cloudy}{https://gitlab.nublado.org/cloudy/cloudy}} \citep{cloudy2023}. The dependence of cooling functions on metallicity, density, and temperature is taken into account using a trilinear interpolation of the cooling curves between the tabulated values. For the \isothermal\, and \coolingflow\, models, we assume a constant metallicity of the CGM gas with values $Z = 0.5 Z_{\rm \odot}$ and $Z = 0.1 Z_{\rm \odot}$ respectively. For the \isentropic\, model, we assume a metallicity profile \citep[following][]{Faerman2020} for the CGM, $Z = Z_0 \big[1+(r/r_z)^2\big]^{-1/2}$ where $Z_0 = Z_{\rm \odot}$ is the Galactic metallicity and $r_z = 90$ kpc is the metallicty length scale. The transport of metals in CGM is dominated by the dynamical evolution of gas compared to diffusion due to turbulence \citep{Shchekinov2002}, hence we trace the metallicity of gas by treating it as a scalar field which follows the advection equation. The gas injected into the CGM due to supernova explosions or AGN is assumed to be at solar metallicity.

\subsection{Mass and energy injections}
\label{subsec:injection}
We consider either multiple supernova explosions or accretion on to central supermassive black holes as drivers of the galactic outflows. We set the radius of the energy injection region, $r_{\rm inj}$, at 200 pc (400 pc for \coolingflow), consistent with the typical star-forming regions in starburst galaxies, for e.g.,~M82 \citep{OConnell1978}, NGC 253 \citep{Mills2021} and the central molecular zone of our own Milky Way \citep{Henshaw2022}. For a typical galaxy, the galactic wind develops a bi-conical shape being collimated by the dense interstellar medium (ISM) in the galactic disk. The collimation of the wind depends on the rate of supernova explosions and the density distribution in the ISM gas. Since we aim to study the interaction between a galactic wind and the CGM we ignore the detailed modeling of the wind collimation during the launch of the wind and assume it is collimated to an opening angle. For all our simulations we choose the half opening angle $\theta_{\rm inj} = \pi/4$. We find that the choice of the injection region and the opening angle has little effect on the emission properties of the interaction zone as the wind quickly develops a shock that is almost spherical in nature. 
\begin{table}
    \caption{Energy injection models}
    \renewcommand{\arraystretch}{2.5}
    \label{tab:injmodel}
    \begin{tabular}{|p{2cm}|p{2.5cm}|p{2cm}|}
    \hline
    \hline
    \makecell{Injection model} & \makecell{Description} & \makecell{Total energy}\\
    \hline
    \makecell{SF, $10^7$ M$_{\odot}$ \\ instantaneous} & \makecell{$10^7$ \msun stars form \\ instantaneously} & \makecell{$1.3 \times 10^{56}$ erg}\\
    \makecell{SF, $10^8$ M$_{\odot}$ \\ instantaneous} & \makecell{$10^8$ \msun stars form \\ instantaneously} & \makecell{$1.3 \times 10^{57}$ erg}\\
    \makecell{SF, $2$ M$_{\odot}$ \\ yr$^{-1}$ continuous} & \makecell{constant SFR = $2$  M$_{\odot}$ \\ yr$^{-1}$ for 50 Myr} & \makecell{$9.6 \times 10^{56}$ erg} \\
    \makecell{SF, $10^9$ M$_{\odot}$ \\ instantaneous} & \makecell{$10^9$ \msun stars form\\ instantaneously} & \makecell{$1.3 \times 10^{58}$ erg}\\
    \makecell {AGN \\ instantaneous} & \makecell{AGN burst\\ over 1 Myr} & \makecell{$1.3 \times10^{58}$ erg}\\
    \makecell {AGN \\ continuous} &  \makecell{AGN burst \\over 40 Myr} & \makecell{$1.3 \times 10^{58}$ erg}\\
    \hline
    \end{tabular}
\end{table}

\begin{figure}
    \centering
    \includegraphics[width=0.5\textwidth]{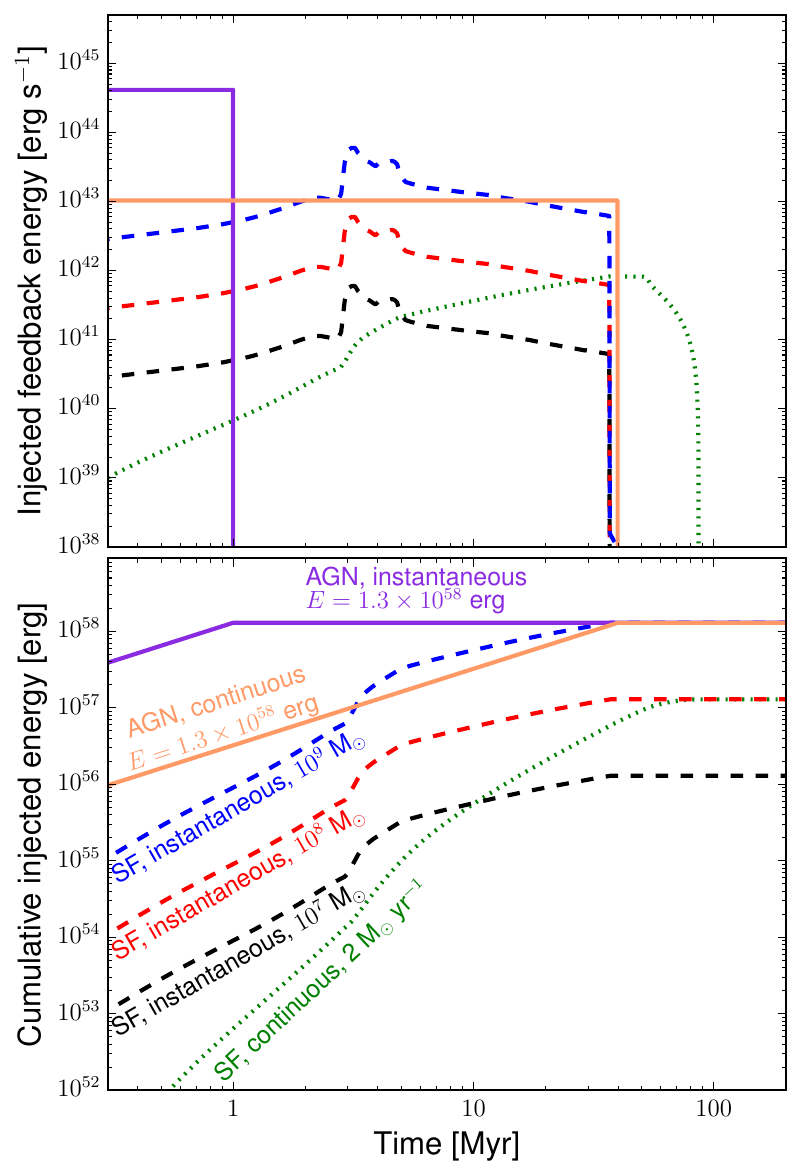}%
    \caption{Different feedback energy injection rates considered in our simulations. The upper panel shows the instantaneous rate while the lower panel shows the cumulative injected energy.}
    \label{fig:inj_param}
\end{figure}
\begin{figure*}
    \includegraphics[width= \textwidth]{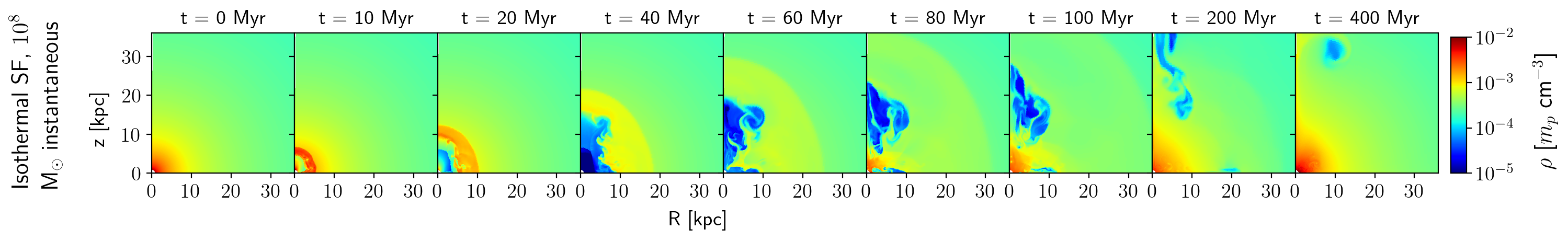}
    \caption{Density distribution versus time of an initially \isothermal\, CGM with feedback energy injection corresponding to an instantaneous SF event of $10^8\,{\rm M}_\odot$. Energy injection occurs at $t<40$ Myr (see Fig.~\ref{fig:inj_param}). Note the progress of the shock wave at a velocity of $\approx 500$ km s$^{-1}$}
    \label{fig:1denmovie}
\end{figure*}

\begin{figure*}
    \includegraphics[width= \textwidth]{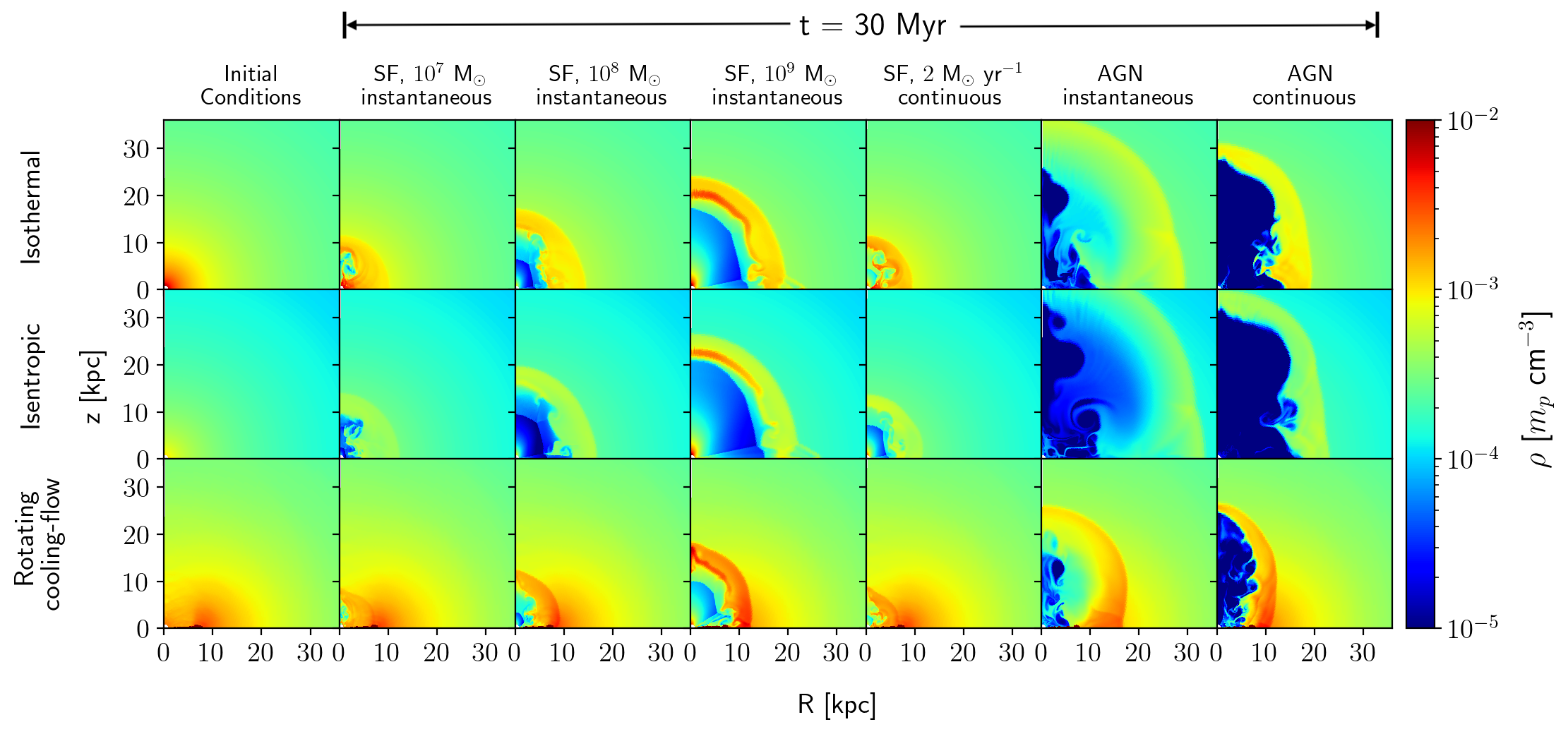}
    \caption{Density distribution of the CGM 30 Myr after the onset of the feedback event, for different initial conditions (different rows)  and different feedback energy injection rates (different columns). The left-most column plots the initial density maps at $t = 0$ Myr. Note the strong sensitivity of the resulting density profile to the initial CGM conditions.}
    \label{fig:2denmovie}
\end{figure*}

\begin{figure*}
    \includegraphics[width= \textwidth]{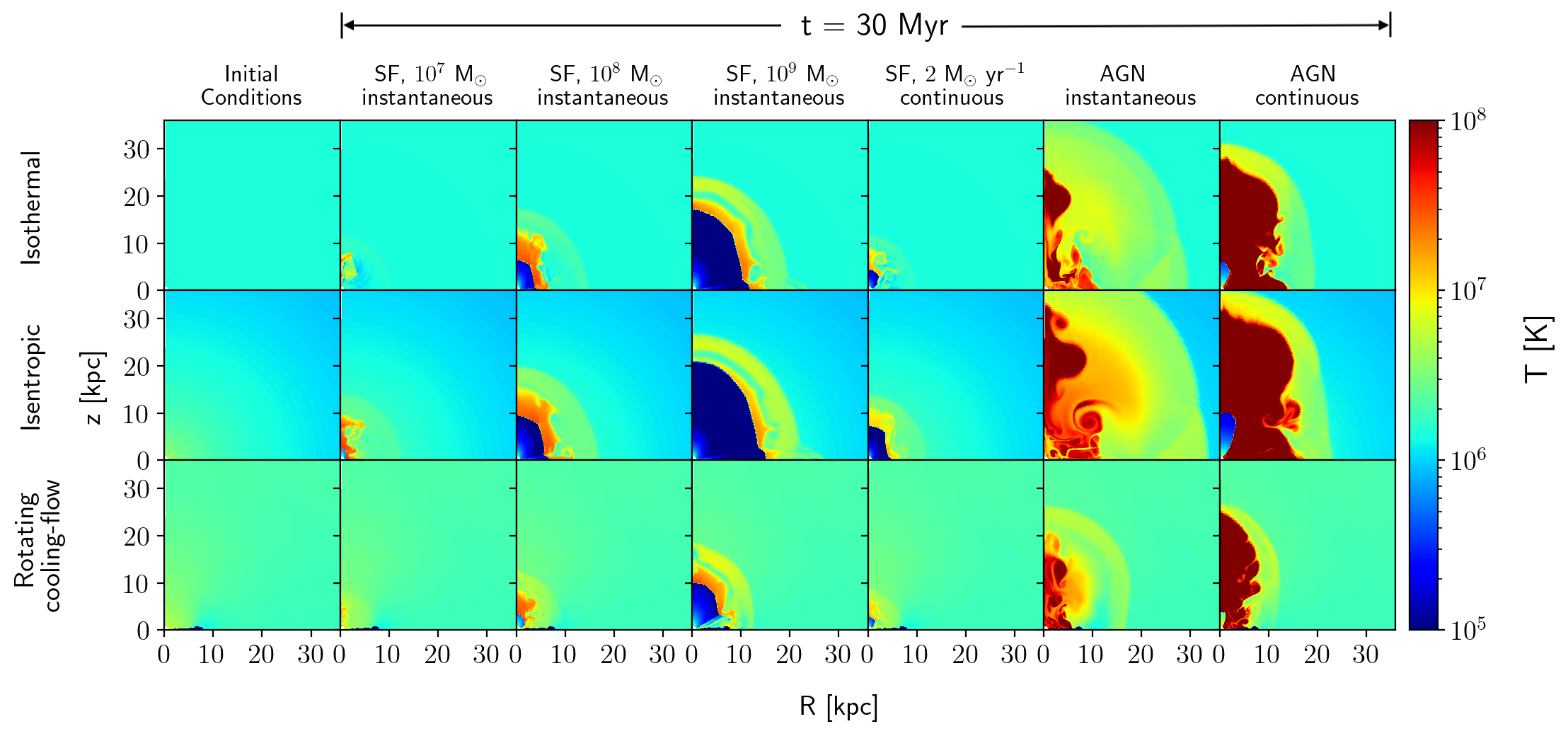}
    \caption{Temperature distribution of the CGM 30 Myr after the onset of the feedback event, for different initial conditions (different rows)  and different feedback energy injection rates (different columns). The left-most column plots the initial temperature maps at $t = 0$ Myr.}
    \label{fig:2tempmovie}
\end{figure*}

\subsection{SNe driven outflow}
Our simulation box starts from $r_{\rm inj}$ that we assume to be the sonic radius of the SNe-driven outflowing gas. Hence, the velocity of the gas at $r_{\rm inj}$ is kept fixed at the velocity of sound at that radius. From the energy conservation equation, we find that the velocity of sound at $r_{\rm inj}$ is half of the free wind velocity, $c_{\rm s} = 0.5 \, v_{\rm fw} = 0.5 \, \sqrt{2L/\dot{M}}$ \citep{Mathews1971, Chevalier1985}. We study two types of scenarios, one in which the stars are formed instantaneously (starburst event) and the other, where the stars form gradually at a constant rate. We calculated the corresponding rate of mechanical energy ($L$) and mass injection ($\dot{M}$) using \textsc{starburst99}\footnote{\hyperlink{https://www.stsci.edu/science/starburst99/docs/default.htm}{https://www.stsci.edu/science/starburst99/docs/default.htm}} model \citep{Leitherer1999} assuming a Kroupa IMF with two exponents 1.3 and 2.3 in the stellar mass range [0.1-0.5] and [0.5-100] \msun~respectively. The boundary value of density ($\rho_b$) and the corresponding pressure ($p_b$) at the sonic radius is calculated from the mass conservation equation
\begin{eqnarray}
    \rho_b & = & \frac{\dot{M}}{\Omega r^2 c_s} \\ \nonumber
    p_b & = & \rho c_s^2/\gamma = \frac{\dot{M}c_s}{\Omega r^2 \gamma}
\end{eqnarray}
 where $\Omega = 4\pi(1-\cos \theta_{\rm inj})$ is the solid angle subtended by the injection cone with $\theta_{\rm inj} = \pi/4$.
 
Figure \ref{fig:inj_param} illustrates the various feedback models implemented in this study. The upper panel displays the rate of energy injection, while the lower panel depicts the total energy injected over time. In each case, energy injection continues beyond the termination of star formation episode as the population of massive wind-driving stars fades away.

\subsection{AGN driven outflow}
We study the effect of an AGN burst, occurring either instantaneously, $t_d =  1$ Myr, or over a time period of $t_d =  40$ Myr (assuming a quasi-steady AGN). For AGN feedback, the wind velocity is assumed to be $v_{\rm agn} = 0.05c$ at $r_{\rm inj}$ where $c$ is the speed of light. The corresponding mass injection rate is given by $\dot{M} = 2\dot{E}_{\rm AGN}/v_{\rm agn}^2$ where $\dot{E}_{\rm AGN} = E_{\rm AGN}/t_d$ . We assume $E_{\rm AGN} = 1.3\times 10^{58}$ erg, corresponding to an AGN power of $\approx 4\times 10^{44}$ \ergps  for $t_d = 1$ Myr and $\approx 10^{43}$ \ergps  for $t_d = 40$ Myr. These AGN powers represent roughly the Eddington luminosity and $2\%$ Eddington luminosity for the Milky Way-type black hole ($M_{\rm BH} \sim 4\times 10^6$ \msun). The boundary value of density ($\rho_{b}$) at injection radius ($r_{\rm inj}$) is calculated from the mass conservation equation
\begin{eqnarray}
    \rho_{b} & = & \frac{\dot{M}}{\Omega r^2 v_{\rm agn}} \\ \nonumber
\end{eqnarray}
 where $\Omega$ is the solid angle subtended by the injection cone with $\theta_{\rm inj} = \pi/4$. The thermal pressure ($p_{b}$) of the AGN-driven wind at $r_{\rm inj}$ is assumed to be equal to the initial ambient pressure at that radius.

We summarize our models in Table \ref{tab:injmodel}. For all the models, we assume that a fraction, $\epsilon = 30\%$, of the total mechanical energy injected by SNe or AGN  is driving the outflow, with the rest radiated away. This fraction, or heating efficiency, is consistent with \textit{CHANDRA} observations of the M82 wind \citep{Strickland2007} and theoretical estimates from numerical simulations \citep{Vasiliev2015, Fielding2018}.

\section{Results}
\label{sec:results}
We now present our simulation results on how the different CGM models respond to the energy injections and the impacts on the X-ray emissions.
 
\subsection{Thermodynamic variables}
\label{subsec:rhoT}
As a first representative example, Figure \ref{fig:1denmovie} shows the time evolution of the density distribution for the initially \isothermal\, CGM model and energy injection method, \eight. Although the stars form in a burst, the energy is injected via stellar winds over time scales corresponding to the lifetime of the stars, as can be seen in Figure \ref{fig:inj_param}. Since the overall energy injection rate from such a starburst is almost constant for the first $\simeq 40$ Myr, the impact of the outflow is very similar to the classical stellar wind solutions \citep{Castor1975, Weaver1977, Silich2004}. It shows the typical structure of a stellar wind with the free wind in the inner region (with adiabatically decreasing density) surrounded by a shocked wind, and then a higher-density shocked CGM gas region. The shocked wind region is created by a strong termination/reverse shock and has a density jump of $4$ compared to the free wind, and therefore, is not often visible in X-ray. On the other hand, even though the density and temperature jump across the forward shock is much lower than that of the reverse shock, the shocked CGM density is high enough to shine in X-rays. Note that the shape of the bubble is not perfectly spherical since we inject energy within a cone with a half opening angle of $\pi/4$. This introduces a small asymmetry in the shock structure. As we will see later in section \ref{sec:analytical}, the propagation of the shock can still be approximately calculated assuming a spherical expansion. 

As noted, for the instantaneous starburst injection methods (\seven, \eight, and \nine), the mechanical luminosity due to stellar winds and SNe declines at around $40$ Myr. Therefore, after a wind crossing time ($\sim \xi/v_{\rm fw}$, where $\xi$ is the distance along the wind cone) the free wind region disappears. On the other hand, the forward shock keeps on moving forward through the CGM. It gradually weakens and ultimately becomes an acoustic wave moving through the diffuse CGM. Shortly after the energy injection rate declines,  the high-density shocked CGM gas loses its ram pressure support and falls inward toward the galactic center giving rise to a complex density and velocity structure. After a long time ($\sim 500$ Myr), the CGM distribution attains an equilibrium configuration. However, in the isothermal and isentropic models, this equilibrium differs from the initial states because we do not include the continuous feedback that maintains the initial states.

Figure \ref{fig:2denmovie} presents a comparative analysis of the density distribution of CGM gas for each of our simulations at a particular time after the onset of the feedback event. The first column represents the initial CGM distribution in each case and all other columns represent the corresponding density distribution at 30 Myr. An under-dense bubble form in the simulations with high injection rates, \nine, \agninst and \agncont, while bubbles do not form in the \seven simulations. In the \eight and \cont a bubble is more apparent in the \isentropic\, initial conditions, which has the lowest central density of the three initial condition models we consider.  Also, an initially higher density CGM will have a smaller forward shock radius at a certain time, as is seen in the \coolingflow\, CGM simulations, since the forward shock velocity decreases with increasing pre-shocked gas density. For continuous SF injection (\cont), initially the mechanical power from the stars grows with time and gradually saturates to a fixed value (see fig \ref{fig:inj_param}). Hence at 30 Myr, when the energy injection rate is still in its growing phase, the outer shock traverses a shorter distance compared to \eight case. We conclude that for an equivalent mass of stars formed, a burst will always push the forward shock further into the CGM compared to more gradual continuous star-formation. 

For the instantaneous AGN feedback model, the wind attains a slightly different internal shock structure compared to the instantaneous SF-driven feedback because the energy is released at once. For such a burst of energy, the dynamics of the shock follow the blast wave-like solution \citep{Sedov1946, Taylor1950}, rather than the wind-like solution \citep{Weaver1977} for a given total energy. At a given time, the outer shock reaches much further into the CGM in \agninst \hspace{0.05cm} compared to starburst events due to the higher energy injection rate of AGN. This is also the reason why the density of the AGN shock appears to be lower than the wind-like case since the density falls off far away from the galaxy.

Figure \ref{fig:2tempmovie} compares the temperature distribution of CGM gas at 30 Myr for each of our simulations. The hottest part of the outflowing gas is the shocked wind region (due to the high Mach number of the reverse shock), which surrounds a cooler free-wind region most apparent in the SF simulations.  The temperature of the shocked wind region in the AGN-driven outflow is $ \gtrsim 10^8$ K which is at least an order of magnitude higher than the star-burst-driven outflow, mainly because of the high specific energy of the AGN driven wind. However, this shocked wind region has a low gas density ($\rho < 10^{-4} m_{\rm p}$ cm$^{-3}$) and thus does not contribute much to the X-ray emission. The dominant contribution to the X-ray emission comes from the surrounding higher density shocked CGM region, which has typical temperatures ranging from a few times $10^6$ K to $10^7$ K.

\begin{figure*}
    \centering
    \includegraphics[width=\textwidth]{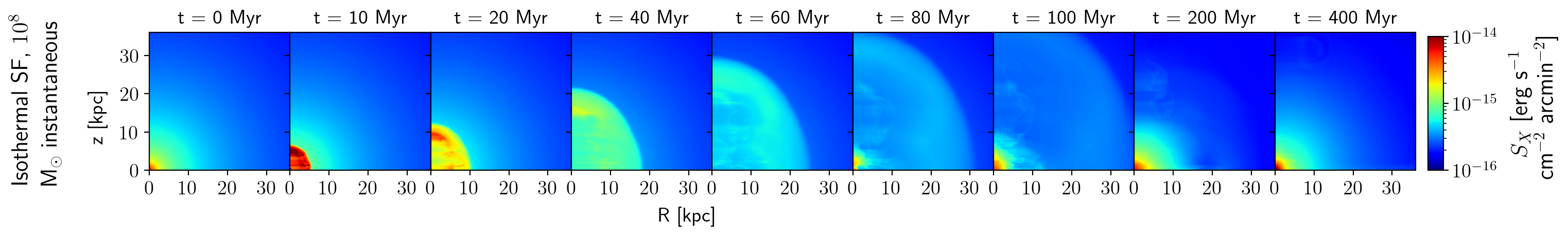}
    \caption{Soft X-ray (0.5-2 keV) surface brightness distribution for \isothermal\, initial conditions, at different times after an instantaneous $10^8 \, {\rm M}_{\rm \odot}$ star formation event. Note the expanding central bright X-ray region due to the shock induced by the outburst.}
    \label{fig:1xraymovie}
\end{figure*}

\begin{figure*}
    \centering
    \includegraphics[width=\textwidth]{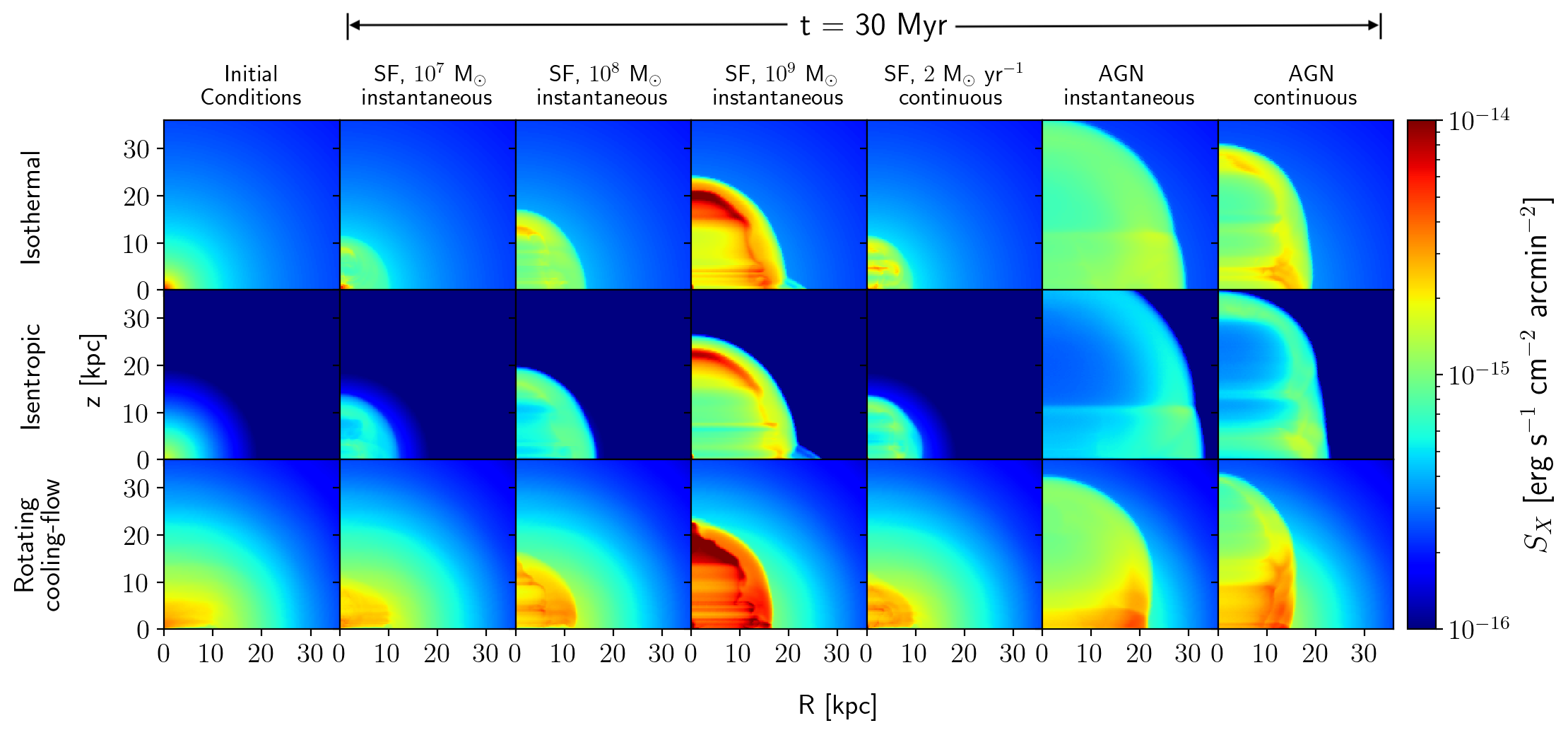}
    \caption{Soft ($0.5-2$ keV) X-ray surface brightness distribution of the CGM 30 Myr after the onset of the feedback event, for different initial conditions (different rows) and different feedback energy injection rates (different columns). The left-most column plots the initial density maps at $t = 0$ Myr.}
    \label{fig:2xraymovie}
\end{figure*}
 
\begin{figure*}
  \centering
  \includegraphics[width=0.9\textwidth]{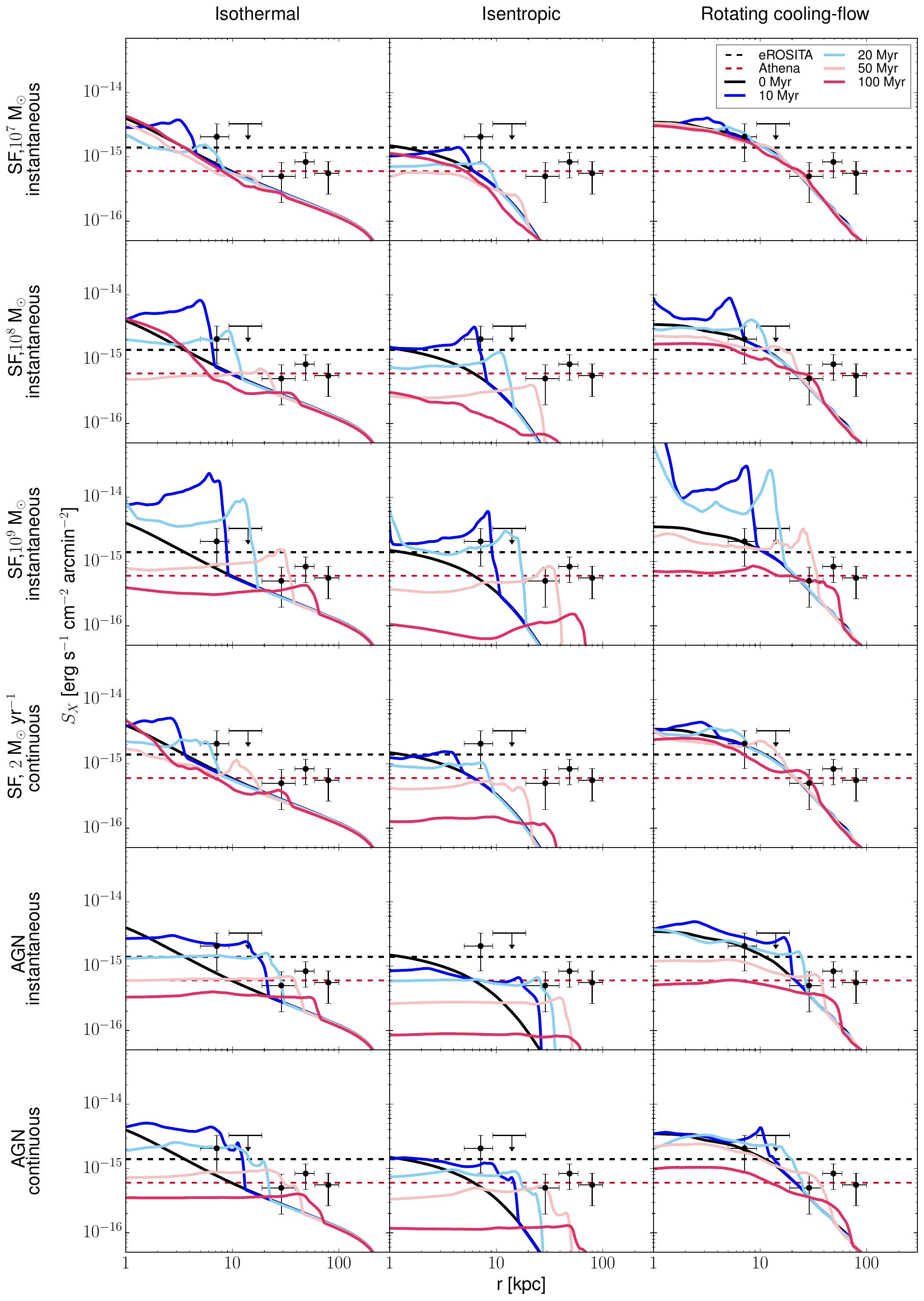}
  \caption{Time evolution of the projected X-ray radial intensity profile for different initial conditions \isothermal, \isentropic, and \coolingflow\, CGM models in case of different feedback models. The black and red horizontal dashed line shows the diffuse X-ray flux ($0.5 - 2$ keV) detection limit of the \erosita \citep{Clerc2018} and \athena \citep{Barret2020} telescopes respectively. The points with error bars represent the stacked surface brightness profile of star-forming galaxies, with stellar mass similar to the Milky Way, from \citet{Chadayammuri2022}.}
  \label{fig:1d-xray-profiles}
\end{figure*}

\subsection{X-ray emission}
\label{subsec:initial-xsb}
X-ray emission from the volume filling hot ($T \sim 10^6$ K) phase is an excellent probe of the CGM gas distribution and its interaction with the outflow. The soft X-ray ($0.5-2.0$ keV) emissivity, $\Lambda_X (n_H, T, Z)$, is calculated from \textsc{cloudy-v22} \citep{cloudy2023} for the given density, metallicity, and temperature of the plasma. Figure \ref{fig:emis_solar} shows the emissivity (per $n_H^2$) for Solar metallicity and assuming the UV background radiation of \cite{Haardt2012ApJ}. 

For a better comparison with the observational results or applicability of our results to future observations, we produce the X-ray surface brightness of the simulated cases. Numerically, the brightness is calculated using \textsc{pass} \footnote{\hyperlink{https://github.com/kcsarkar/PASS-EOV}{https://github.com/kcsarkar/PASS-EOV}} \citep{Sarkar2017} and assuming the above noted emissivities. We consider a 3D sphere with a radius of $r_{\rm vir} (= 258$ kpc) to compute the projected X-ray surface brightness maps. We project the brightness on the $y-z$ plane and integrate the X-ray volume emissivity along the $x$-direction (in Cartesian coordinate terms) following
\begin{eqnarray}\label{eqn:xray}
     S_X(y, z) &=& \frac{1}{4\pi \times 1.2\times 10^7} \int n_H^2 \Lambda_X (n_H, T, Z) \, dx \nonumber \\ 
     && \hspace{2cm} {\rm erg \, s^{-1} \, cm^{-2} \, arcmin^{-2}} ~.
\end{eqnarray}

\subsection{X-ray projections maps}
\label{subsec:feedback-xsb}
\subsubsection{2D surface brightness profiles}
Figure \ref{fig:1xraymovie} shows the evolution of the projected X-ray surface brightness map for one representative case with the \isothermal\, CGM model and \eight. The bright spherical region that grows with time from the galactic center marks the extent of the forward shock. As we see in the density maps, the injected energy, be it wind-like or burst-like, creates a low-density bubble surrounded by a dense shell region. Since the X-ray emissivity is proportional to density squared ($S_X \propto n_H^2 $, see eq. \ref{eqn:xray}), most of the contribution to the X-ray surface brightness originates from the dense and shock-heated gas near the outer shock region and produces the limb-brightened features as seen in the figure. As we mentioned earlier, $\approx 40$ Myr after the onset of the star formation, the energy injection rate declines substantially. As a result the forward shock gradually becomes weak giving rise to a fading X-ray bubble. But the shocked gas falls towards the galactic center due to the loss of ram pressure support. Due to this, a rebrightening of the inner CGM occurs ($r < 10$ kpc) after 80 Myr and onwards, as seen in Figure \ref{fig:1xraymovie}.

Figure \ref{fig:2xraymovie} shows the X-ray surface brightness map of all our simulations at 30 Myr. The strong feedback simulations (\nine, \agninst and \agncont) show X-ray surface brightness maps which greatly differ from the initial conditions (with most of the brightness coming from the shell of the shocked CGM gas) while in the weaker simulations only mild differences are apparent.

\subsubsection{Radial X-ray profiles and comparison with observations}
\label{subsec:xray-profiles}
Figure \ref{fig:1d-xray-profiles} shows the spherically averaged surface brightness profiles for all the simulations we performed. The black solid curves in all the panels show the initial angle averaged X-ray radial profiles. The other solid curves represent the X-ray profiles at different subsequent times after the onset of energy injection. The black points with error bars represent the \erosita stacked surface brightness profile of star-forming galaxies, with stellar masses similar to the Milky Way \citep{Chadayammuri2022}. The black and red dashed horizontal lines denote the observational detection limit for \erosita and \athena telescopes respectively. The detection limit of \erosita in $0.5-2$ keV range for extended sources is calculated from a synthetic simulation of the extra-galactic sky \citep{Clerc2018, Merloni2012} taking an exposure time $t_{\rm exp} \sim 20$ ks. The \athena observation for extended objects is likely background limited (A.~Bamba, priv.~comm), so we calculate the detection limit by integrating the total (photon+particle) background model of the Wide Field Imager within $0.5-2$ keV energy range and then dividing by the effective area of $1.4 \, \rm m^2$ \citep[for $E \sim 1$ keV;][]{Barret2020}  of the telescope. Comparing with the detection limit of \erosita and \athena we notice that the initial CGM profiles can be detected in X-rays up to $10-20$ kpc from the galactic center if CGM follows an \isothermal\, or \coolingflow\, model. The X-ray radial profile for \isentropic\, CGM falls below the detection limit after $\sim 5$ kpc. However, this limit for the \isentropic\, CGM can be extended up to 10-20 kpc in the presence of feedback. 

For strong feedback models (\nine, \agninst and \agncont) the main feature in the radial X-ray profiles are the peaks that mark the positions of the outer shocks. Internal to the peak positions the X-ray profiles are shallower compared to the initial profile. The peaks arise due to the dense hot gas near the forward shock as already discussed, while the X-ray brightness of the free wind region is diminished due to the low gas density. The presence of the X-ray bright shocked-CGM just outside the X-ray cavity creates an edge-brightened surface brightness profile due to projection effects. Identifying such a flat X-ray emission profile followed by a peak would provide evidence for a recent strong feedback event.   The three outer points from the \cite{Chadayammuri2022} may provide some evidence for such a flattening. 
In contrast, in models where feedback is weak (\seven ), the X-ray emission profile remains close to its initial shape.

The height of the peak depends on various properties of the feedback model, including the total energy injected and whether the injection is instantaneous or continuous. Though the total injected energy is almost the same for \eight and \cont, the shocks are stronger for instantaneous injection giving rise to higher peaks in the X-ray intensity profiles. The height of the peak also depends on the initial CGM density, since initially higher density CGM (\isothermal\, and \coolingflow\, CGMs) are brighter than the low density \isentropic\, model. Thus, for a given energy injection rate, the \isentropic\, CGM simulation exhibits more prominent peaks and more substantial flattening of the inner X-ray surface brightness, compared to the \coolingflow\, CGM.

We conclude that the X-ray surface brightness due to CGM-outflow interaction can increase up to 10 times its initial value for our studied feedback parameters and can be detected by telescopes like \erosita and \athena. We also notice that while the \isentropic\, and \coolingflow\, CGMs can be brightened up by some reasonable energy injection to match with the \erosita data, the \isentropic\, model falls significantly short of producing the observed level of X-ray brightness at $r\gtrsim 50$ kpc.

\begin{figure*}
    \centering
    \includegraphics[width=0.8\textwidth]{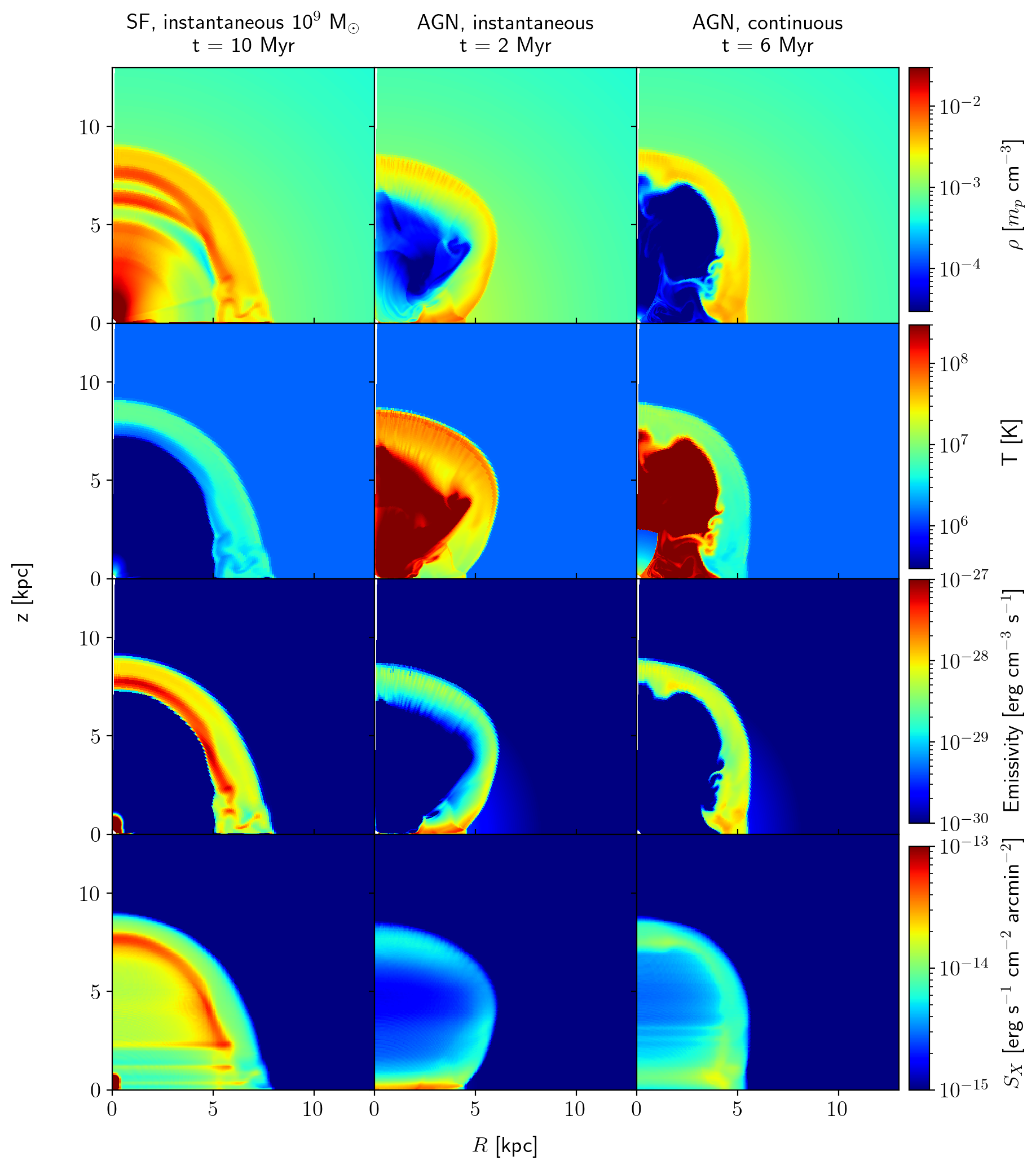}
    \includegraphics[width=0.8\textwidth]{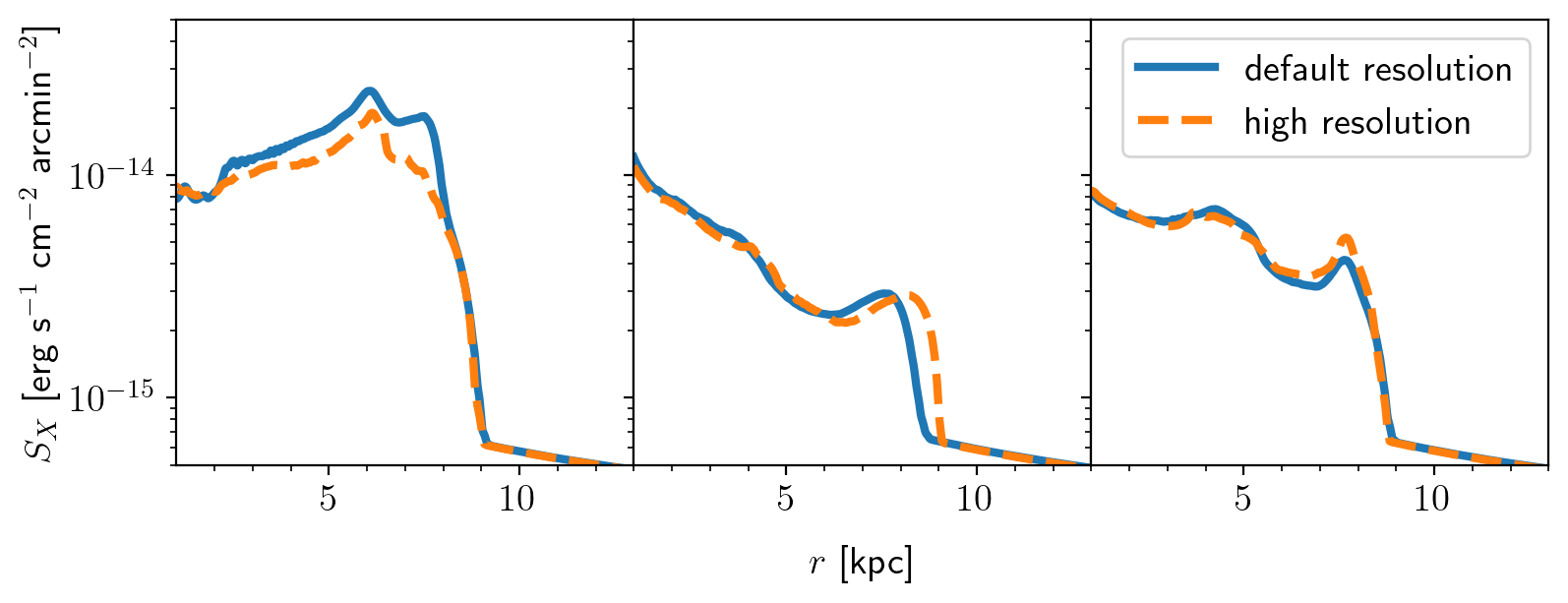}
    \caption{Comparison between star-formation and AGN-driven shock cases. The lowest panel shows the $\theta$-averaged radial surface brightness profiles for the simulations. The dashed lines show the same but for the high-resolution simulations (for details see Section \ref{sec:res_test}).}
    \label{fig:SX-at-given-r}
\end{figure*}

\subsection{Star-formation versus AGN}
At this point, it is difficult to predict the nature of feedback just by looking at the X-ray intensity profiles. However, some indications of such a distinction may be found from a more careful look at the surface brightness profiles. Figure \ref{fig:SX-at-given-r} shows the density, temperature, and X-ray emission properties between the \nine, \agninst, and \agncont simulations where the total injected energies are the same. Unlike the earlier subsection where we compared the brightness maps and profiles at a given time, this figure compares them when they reach a given radius. This is because at a given time, the shock radius in the \agninst case is much larger and the local CGM density is entirely different. Therefore, to understand the impact of the different energy injection methods alone, we compare the surface brightness when the shocks reach a given radius of $9$ kpc. 

The top panel of figure \ref{fig:SX-at-given-r} shows the density snapshots for the different energy injection methods. In the \agninst case, the shock can be treated as a blast wave; hence, most of the swept-up CGM lies closer to the shock front, whereas, for the continuous case, the swept-up matter is spread out in a thicker shell behind the shock. Such a thick shell is also expected to be seen in any continuously powered bubbles where radiative cooling is negligible \citep{Castor1975, Weaver1977}. However, the main difference between the \agninst case and the \agncont case is that the former has already injected the entire $1.3\times 10^{58}$ erg energy and hence the shock velocity is much higher. With a high shock velocity, the shock temperature turns out to be $\sim 10^8$ K where the X-ray emissivity is about a factor of $2-3$ lower than the value at $\sim 10^6 - 10^7$ K (see Figure \ref{fig:emis_solar}). This leads to the shocked CGM in the \agncont case having slightly higher soft X-ray emissivity than the instantaneous case, as seen in the third row of the figure.

It would be only justified to compare the X-ray emission from the \nine to the \agncont case since they have a similar injection power. Note that even though the SF case ($t=10$ Myr) injects a factor of $4$ more energy than the AGN-continuous case ($t=6$ Myr), the actual energy driving the shock is very similar for the following reasons. The SF-driven wind is carried out from the galactic center to the forward shock ($r\sim 9$ kpc) at a velocity of $v_{fw} = \sqrt{2 L/\dot{M}} \approx 1800$ \kmps, i.e over a delay time of $t_{\rm delay} = r/v_{fw} \approx 5$ Myr. The same delay time for the AGN case is $t_{\rm delay} \approx 0.6$ Myr, assuming $v_{\rm agn} = 0.05 c$. 
Therefore, even though a factor of 4 more energy is emitted by the SF at $t=10$ Myr, almost half of the energy (i.e. $E(10\mbox{Myr})-E(5\mbox{Myr})$) is not used to push the shock. Therefore, the actual energy powering the SF shock and the AGN-continuous shock is very similar.

The main difference between the \nine and the \agncont case is the presence of a dense, warm shell of $T\sim 10^6$ K at the contact discontinuity. This shell appears due to the mass ejected by the SNe remnants. For the case of \nine simulations, the total SNe-injected mass by $10$ Myr is almost the same as the total CGM mass within the central $9$ kpc. Now, while the CGM mass is swept up into the shocked CGM shell, the SNe ejected material can not cross the contact discontinuity (due to the pressure uniformity within the shocked wind and shocked CGM) and hence stacks up at the contact discontinuity. Although this stacked gas cools and fragments over time (see figure \ref{fig:res_check}), there is enough material with temperature $\sim$ few $\times\, 10^6$ K that can emit X-ray. Given that this gas has a higher density than the shocked CGM, the X-ray emission from the contact discontinuity dominates the total surface brightness as can be seen in figure \ref{fig:SX-at-given-r}. However, we note that such a bright X-ray emission from the contact discontinuity is only seen when the SNe-injected mass is comparable to the swept-up CGM mass and is dependent on the star formation rate. For \eight case, we do not notice such a large enhancement of X-ray brightness. The enhancement is rather less than a factor of 2. In such mid to low-range SF cases, the X-ray surface brightness is not much different from the corresponding AGN cases with similar energy. The indifference of SF and AGN in these cases arises from the fact that the shock velocity depends only on the power of the injection, the background density, and the time, i.e. $r_s \approx (L t^3/\rho)^{1/5}$ (see further discussion below). Therefore, for a given power (or total energy), the shock velocity, the shock temperature, and the X-ray emission are independent of the type of the wind itself. A tale-tell sign for the X-ray emission from contact discontinuity is the double-peaked X-ray emission near the shock (see bottom-left panel of figure \ref{fig:SX-at-given-r}). The peaks are located within $\sim 20\%$ of each other, as expected from the stellar wind bubble models \citep{Weaver1977}.

\section{Comparison of simulation results with analytical calculations}
\label{sec:analytical}

We have found that most of the X-ray surface brightness comes from the dense and hot shocked gas region right behind the outer shock. The thickness of this highly X-ray emitting shell is almost $50$\% of the shock radius. To estimate the X-ray surface brightness for outflow-CGM interaction, one needs to know the density-temperature structures and metallicity. It is possible to provide an analytical estimate for the position of the outer shock assuming a stellar wind model for the outflow \citep{Weaver1977} but for a power-law density profile. We study the analytical solution for our model \eight and compare it with our simulation results. As we see in Figure \ref{fig:inj_param}, upper panel, even though the energy injection rate is a function of time, for the chosen feedback model, it can be approximated by a roughly constant injection rate of $L = 10^{42}$ erg s$^{-1}$ for $\approx 40$ Myr.

\begin{figure*}
\hspace{-0.3cm}
\includegraphics[width=\textwidth]{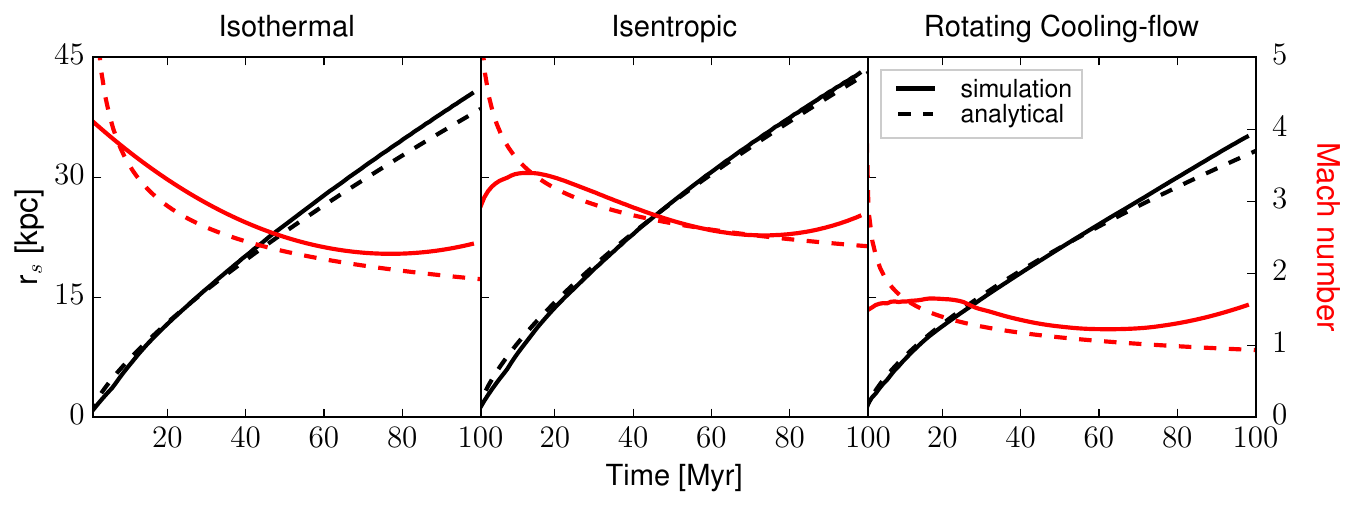}
\caption{The black and red solid curves show the polar angle averaged outer shock position and the Mach number from the simulation for instantaneous star formation of $10^8 \, {\rm M}_{\odot}$ stars. The black and red dashed curves are the analytical estimates of the shock radius and Mach number if the outflow follows a stellar wind solution with an ambient power-law density profile (equation \ref{eq:shock_rad}) similar to the respective CGM model. The Figure shows that the analytical estimates of the outer shock positions are in good agreement with the same obtained from the simulation for all three CGM models.The analytically estimated Mach numbers are not good estimates for the same obtained from the simulation.} \label{fig:shock_rad}
\end{figure*}

\begin{figure*}
\includegraphics[width=\textwidth]{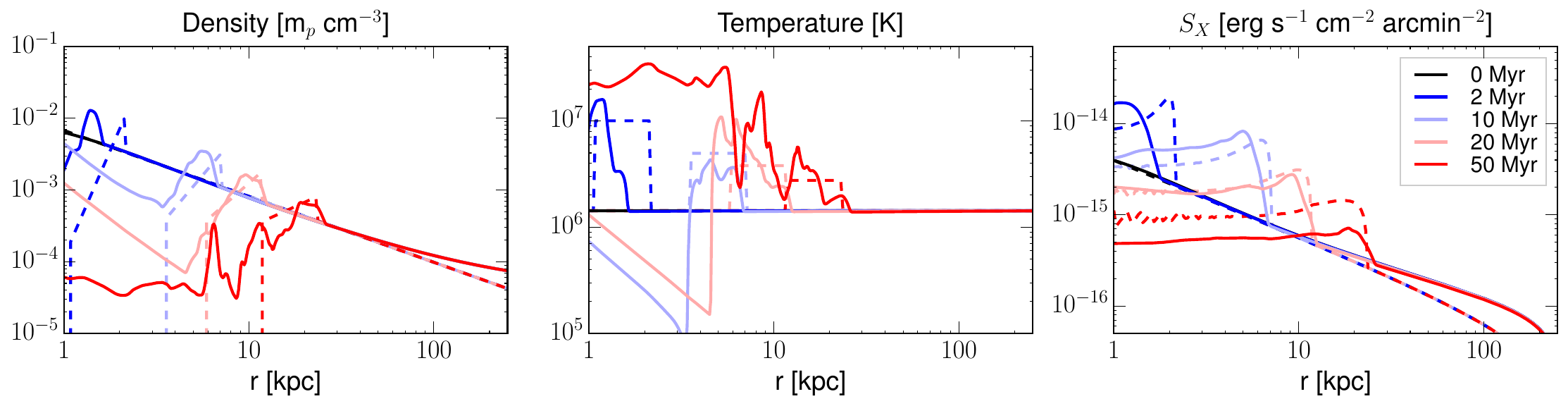}
\caption{Comparison of analytic spherical shock propagation calculation (dashed curves, see section \ref{sec:analytical}) with polar-angle averaged values in the simulations (solid curves). The three panels show the density, temperature, and radial X-ray surface brightness distributions for different times, assuming an initial \isothermal\, CGM model and an instantaneous $10^8 \, {\rm M}_{\odot}$ star-formation event.}\label{fig:ana-com-simu}
\end{figure*}

We model the outflowing gas originating from the galactic center using the \citet{Weaver1977}'s stellar wind solution, which neglects pressure of the ambient medium and hence expected to be applicable when feedback is sufficiently strong. The outer shock position in the stellar wind solution is given as $r_{\rm s} = A (L t^3/\rho)^{1/5}$ where $A \sim 1$, $L$ is the injected mechanical luminosity, $t$ is time and $\rho$ is ambient density. For a power law density profile $\rho(r) = \rho_0 \left(\frac{r}{r_0}\right)^{-\alpha}$ the forward shock position $r_s$ and its velocity $v_s$ can be written as
\begin{eqnarray}
r_s &=&\left[\frac{6.7\: L_{44}}{n_0\: r_{0, {\rm kpc}}^\alpha} \right]^{1/(5-\alpha)}\, t_{\rm Myr}^{3/(5-\alpha)} \quad \mbox{kpc}\nonumber \\ 
v_s & = & 980\left(\frac{3}{5-\alpha}\right) \frac{r_{s, \rm kpc}}{t_{\rm Myr}} \quad \mbox{km s$^{-1}$}
\label{eq:shock_rad}
\end{eqnarray}
where, $L_{44} = L/10^{44}$ \ergps, $n_0 = \rho_0/m_p$, $r_{0,{\rm kpc}} = r_0$/kpc, and $t_{\rm Myr} = t/$Myr.
The \isothermal, \isentropic, and \coolingflow\, models can be well approximated by a power law density profile with $r_0 = 1$ kpc, $n_0 = 6.8\times10^{-3}, 1.3\times10^{-3}, 2\times 10^{-3}$ \pcc and $\alpha = 0.92, 0.6, 0.4$, respectively for the inner 40 kpc region. As an example, for the \isothermal\, model, the outer shock position is given by:
\begin{equation}
    r_s =  11.8 \, {\rm kpc} \left(\frac{L} {\epsilon_{0.3} \times 10^{42}\, {\rm erg \, s^{-1}}}\right)^{0.25} \, \Big(\frac{t}{20\, {\rm Myr}}\Big)^{0.74} \\
    \label{eq:shock_rad_fid}
\end{equation}
where $\epsilon_{0.3} = \epsilon/0.3$ is the heating efficiency factor. In Figure \ref{fig:shock_rad} we compare the outer shock position from the simulation (black solid curve) of the three CGM models for \eight, with our analytical estimate of the outer shock position (black dashed curve) for the same configurations assuming it follows a stellar wind structure according to Equations \ref{eq:shock_rad}. The figure shows that the stellar wind solution is a good estimate for the shock location from the simulation within $\approx$ 40 Myr. At around $40$ Myr, there is a sharp decline in the energy injection rate (see Figure \ref{fig:inj_param}) and the stellar wind solution no longer holds. However, the accuracy of shock position calculation remains within 10\% of the theoretical prediction for up to 100 Myr. 

Once we know the shock radius for each case, we can estimate the shock velocity and the Mach number of the shock, $\mathcal{M}=v_s/c_s$, where $c_s$ is the sound speed of the preshocked CGM at the shock radius. In Figure \ref{fig:shock_rad}, the red solid and dashed curves represent $\mathcal{M}$ calculated from the simulation and analytical model respectively. As mentioned above the \citet{Weaver1977}'s stellar wind solution is strictly valid only for a negligible preshocked gas pressure, i.e. for strong shocks ($\mathcal{M}^2\gg1$). In our simulations, $\mathcal{M}$ can be only slightly larger than unity when feedback is relatively weak (an instantaneous SF of $10^7$ or $10^8$ \msun). At these low feedback energy injection rates the Weaver solution should be modified to account for the finite CGM pressure.

For the \isothermal\, model, the temperature of the CGM is constant throughout but for other models, the temperature profiles can be approximated as a power law profile in the radius range of interest. We calculate the density jump ($\chi$) and temperature of the post-shock gas ($T_2$) following the Rankine-Hugoniot jump conditions for the given Mach number of the shock \citep{Vink2015} as follows:
\begin{eqnarray}
    \chi \equiv \frac{\rho_2}{\rho_1} & = & \frac{(\gamma+1)\mathcal{M}_s^2}{(\gamma-1)\mathcal{M}_s^2+2} \nonumber\\
    \frac{kT_{e,2}}{\mu_e m_p v_1^2} & = & \frac{1}{\gamma \mathcal{M}_s^2}\Big(\frac{\mu}{\mu_e}\Big)+\frac{1}{2}\Big(\frac{\gamma-1}{\gamma}\Big)\Big(1-\frac{1}{\chi^2}\Big) \nonumber \\
    \frac{kT_{i,2}}{\mu_e m_p v_1^2} & = & \frac{1}{\gamma \mathcal{M}_s^2}\Big(\frac{\mu}{\mu_i}\Big)+\frac{1}{2}\Big(\frac{\gamma-1}{\gamma}\Big)\Big(1-\frac{1}{\chi^2}\Big) \nonumber \\
    T_2 & = & (T_{e,2} + T_{i,2})/2
\end{eqnarray}
Here subscripts 1 and 2 indicate the quantities upstream and downstream of the shock respectively and $v$ is the plasma velocity in the frame comoving with the shock. The adiabatic index $\gamma = 5/3$, and $\mu_e = m_e/m_p \approx 1/1836$, $\mu_i \approx 1.27$. As our simulation considers a single fluid approximation, we define the gas temperature in the post-shock region as the average of the ion and electron temperatures.

For a full computation of the X-ray brightness profile, we need a density and temperature profile behind the shock. A detailed analytical model for the post-shock gas profile in a power-law density profile is beyond the scope of this paper. Instead, we make a simplistic assumption that the shocked gas forms a shell with width, $w_r r_s$ in the downstream region of the outer shock, and that the density is zero for $r< w_r r_s$. The value of $w_r$ depends on the exact ambient profile, for example, $w_r \simeq 0.15$ for $\alpha =0$ \citep{Weaver1977} but can be larger for $\alpha>0$. We find that $w_r= 0.5$  matches well with most of our simulations. We model the gas density in the shell with a power law distribution, $\rho(r) = \rho_s (r/r_s)^{-\alpha_s}$, such that the total mass within the outer shock ($M_s$) remains constant. Note that the total mass within the shell is the same as the total mass of the initial CGM within $r_s$ since a strong shock sweeps up almost all the mass inside $r_s$ and compressed it into a shell, while the mass injected by feedback is small except in the \nine simulation. Therefore, the mass within the shell is given by
\begin{eqnarray}
\label{eq:alpha_s}
M_s(<r_s) & = & \int_{(1-w_r)r_s}^{r_s} \rho_s \left(\frac{r}{r_s}\right)^{-\alpha_s} \, 4\pi r^2 dr \nonumber \\
& = & \frac{4 \pi \rho_s r_s^3}{3 - \alpha_s} \left(r_s^{3-\alpha_s} - ((1-w_r)r_s)^{3-\alpha_s}\right) \quad.
\end{eqnarray}

We can now use the known value of the CGM mass $M_s(<r_s)$ to find the value of $\alpha_s$ at that shock radius. As mentioned above, shocks in our simulations are not strong when the feedback injection rate is low (\seven or \eight), so the strong shock approximation is applicable to cases where feedback is stronger.

We find the slope ($\alpha_s$) of the power law density distribution within the shell after numerically solving the above equation. The temperature of the shell is kept constant and equal to the postshock gas temperature. The dashed curves in the left and middle panel of Figure \ref{fig:ana-com-simu} show the analytically obtained density and temperature distribution of the CGM as a function of time for constant energy input (equivalent to the \eight case) in the \isothermal\, CGM model. The solid curves on these two panels show the angle-averaged profiles from our simulation (case, \eight). The density and temperature jump at the shock front from analytical results are in good agreement with the simulation results justifying our use of the analytical formalism to calculate the X-ray surface brightness in this case. We note that the temperature profile obtained from the simulation at 50 Myr shows very high temperature in the inner 10 kpc and does not match with the analytically obtained profile. This occurs because the free wind loses its structure after $\approx 40$ Myr when the energy injection rate drops substantially, and the very diffuse gas in the hot shocked wind region fills the inner region losing ram pressure support. The analytical model only works when there is a constant energy injection rate and hence, is not a good estimate for the profiles after $\approx 40$ Myr. However, the hot gas in the inner 10 kpc region in the 50 Myr profile does not contribute much to the X-ray emission because of its low density. 

Using the X-ray emissivity tables calculated from \textsc{cloudy-v22} (see Figure \ref{fig:emis_solar}), we can now calculate the X-ray surface brightness using our analytic formulae. The right most panel of Figure \ref{fig:ana-com-simu} compares the radial X-ray surface brightness profiles for our analytically solved and computationally obtained CGM distribution as a function of time. Although we cannot accurately model the density and temperature in the free wind and shocked wind region, the X-ray surface brightness calculated from our analytical model is a good estimate of our simulation results. This is because most of the X-ray emission comes from the hot and dense shocked gas region and analytically estimating the shell density and temperature is sufficient to estimate the X-ray emission from the outflow-CGM interaction. 

We note that the analytical model can predict the forward shock position for all the feedback models featuring a constant rate of energy injection, including the AGN-driven winds. The model's performance in calculating the time-dependent X-ray surface brightness varies depending on the specific feedback scenario and initial CGM conditions. For \isothermal\, and \isentropic\, CGM, which can be approximated by a single power-law within the inner 40 kpc region, the analytical model effectively predicts the X-ray emission profiles. In contrast, the \coolingflow\, solution does not conform well to a single power-law profile at the inner 40 kpc due to the density and temperature variation with the polar angle $\theta$, as a result of centrifugal forces in the rotating hot gas (see also \cite{Sormani2018}). Despite this limitation, the analytic model still performs well in calculating the shock position for \coolingflow\, by assuming a single power-law model that follows the CGM distribution along the $z$ direction. Additionally, regardless of the initial CGM profile, the analytical model exhibits lower accuracy in calculating the time-dependent X-ray radial profiles for the \nine feedback model compared to other feedback models. This discrepancy arises because the injected mass by SNe in the \nine model is comparable to the CGM mass enclosed within the forward shock for a period of $\approx 40$ Myr. The significant contribution of the SNe injected mass in X-ray emission leads to a deviation from the assumptions made in the analytical model, resulting in reduced accuracy in predicting the X-ray radial profiles for this specific feedback scenario.

\begin{figure*}
    \centering
     \includegraphics[width=\textwidth]{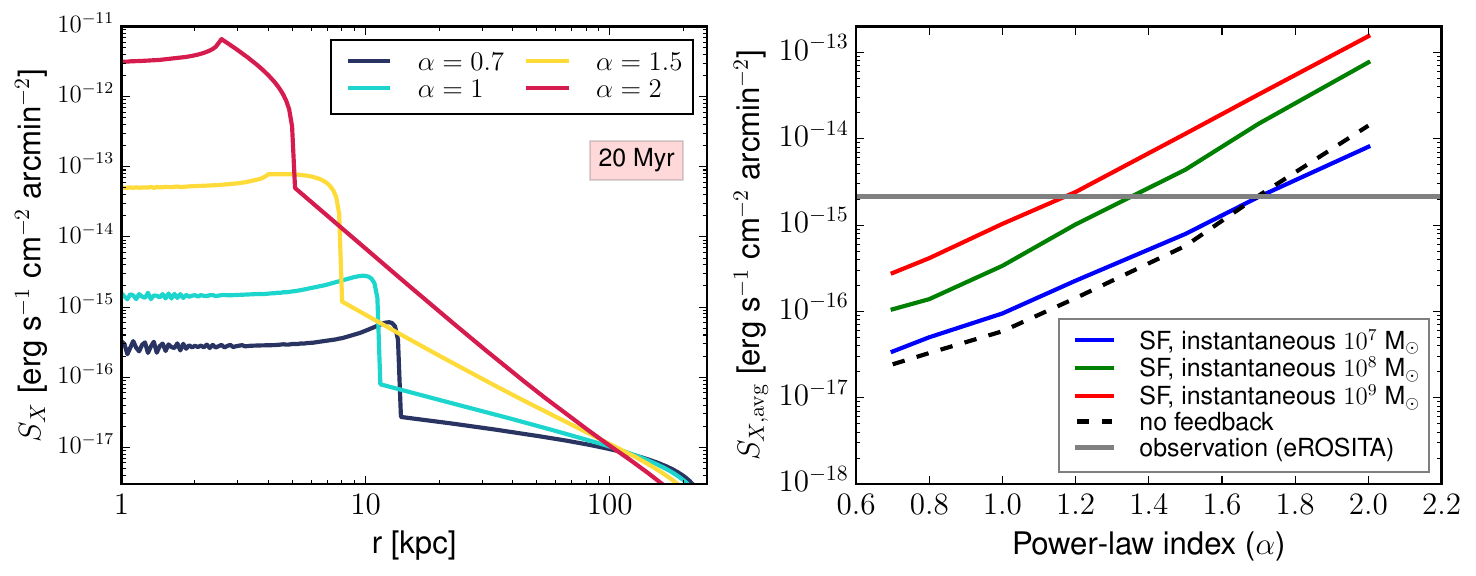}
    \caption{Left panel: Radial profile of X-ray surface brightness at 20 Myr for a CGM power-law profile $\rho(r) = \rho_0 \left(\frac{r}{r_0}\right)^{-\alpha}$ with different $\alpha$ (shown using different line colors) but same total CGM mass ($M_{\rm tot} = 10^{11}$ \msun)  and \eight feedback model. Right panel: Average X-ray surface brightness (calculated within 30 kpc, at 20 Myr) as a function of $\alpha$.  For all the feedback models, shown using different line colors, the correlation between $S_{X, \rm avg}$ and $\alpha$ best fits with an exponential profile. The gray horizontal line shows the average X-ray surface brightness observed in \erosita survey for Milky Way sized galaxies.}
    \label{fig:power-law}
\end{figure*}

\section{Discussion}
\label{sec:discussion}
\subsection{Dependence of X-ray emission on initial CGM distribution}
For this study, we have chosen three different initial CGM models. There are several other models of CGM gas distribution (e.g.,~\citealt{Maller2004, Voit2019}) either obtained from analytical calculations or simulations to explain different observables. The outflow propagates in a different way in each case, for example, it traverses faster in a less dense CGM. The X-ray emission depends on the density, temperature, and shell width behind the forward shock, as noted in section \ref{sec:analytical}. The simulation for each model is computationally expensive but if a CGM model can be approximated by a power law profile, our analytical model can give a good estimate for the actual density and temperature change and subsequent X-ray emission. The left panel of Figure \ref{fig:power-law} shows a comparison of radial X-ray surface brightness profiles for CGM power-law models $\rho(r) = \rho_0 \left(\frac{r}{r_0}\right)^{-\alpha}$ with varying $\alpha$ obtained from our analytical model. All the CGM profiles have the same total CGM mass, $M_{\rm tot} = 10^{11}$ \msun, same CGM temperature ($T = 10^6$ K) and metallicity ($Z = 0.3 Z_\odot$). The profiles are obtained at 20 Myr after the onset of \eight feedback event. We note that the location of the outer shock ($r_s$) is dependent on the slope of the power-law density profile, but, due to normalization, each profile also exhibits a distinct central density ($n_0$). The dependence of outer shock position over $\alpha$ and $n_0$ adheres to the relationship outlined in Equation \ref{eq:shock_rad}. The figure illustrates a strong correlation between the nature of the initial CGM profile and the X-ray emission profiles, indicating that the brightness of the emission varies considerably with the value of $\alpha$, where higher values result in a brighter profile. 

We calculate the average surface brightness ($S_{X, \rm avg}$) within the inner 30 kpc region from the radial surface brightness profile at 20 Myr.
\begin{equation}
    S_{X,\rm avg} = \frac{1}{\pi\cdot (30\,{\rm kpc})^2}\int_0^{30} 2\,\pi r S_{X} dr
\end{equation}
We find that $S_{X, \rm avg}$ increases exponentially with the power-law index ($\alpha$) of the CGM density profile, though we note that $S_{X, \rm avg}$ depends sensitively on the assumed initial temperature, due to the high sensitivity of the emissivity to $T$ at $T\approx10^6$ K (see Figure \ref{fig:emis_solar}). In Figure \ref{fig:power-law}, right panel we show the $S_{X, \rm avg}$ calculated for different feedback models as a function of $\alpha$. The gray horizontal line shows the average X-ray surface brightness observed in \erosita survey \citep{Chadayammuri2022} within the inner 30 kpc of star-forming Milky-Way-sized galaxies. For an observed value of $S_{X, {\rm avg}}$, therefore, this figure shows a degeneracy between the initial CGM density profile and energetics of ongoing feedback activities which can produce the observed X-ray surface brightness. Breaking this degeneracy is possible if we can map the surface brightness profile beyond the shock.

\begin{figure}
    \centering
    \includegraphics[width=0.45\textwidth]{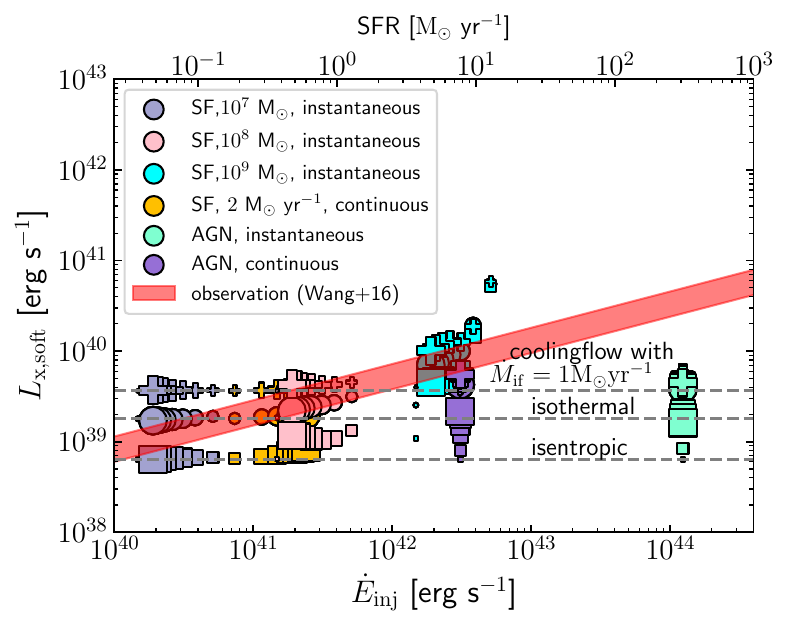}
      \caption{Comparison of total X-ray luminosity ($L_X$), of inner 30 kpc CGM, as a function of
      injected input energy with the \textit{CHANDRA} observation. The red-shaded region shows the best-
      fit relation of $L_X$ vs SFR of a Milky Way mass galaxy, obtained from 52 \textit{CHANDRA}-observed nearby highly inclined disk galaxies by \citet{Wang2016}. The different markers, obtained
      from our simulation, represent the X-ray luminosity for different cases. The shapes of the markers represent different initial CGM conditions (the circles represent \isothermal\,, squares represent \isentropic\, and the plus signs represent the \coolingflow\, model) and the color represents different feedback methods. The size of the markers represents time; a larger size represents a later time (maximum size = 40 Myr). The horizontal gray dashed lines show the total X-ray luminosity of the initial CGM gas distributions we assumed. Our simulation results show a sub-linear relation between the X-ray luminosity and energy injection rate, consistent with the observation. We also note that for low energy injection rate ($\dot{E}_{\rm inj} \lesssim 10^{41}$ erg s$^{-1}$ for \isothermal\, and \isentropic\, and $\dot{E}_{\rm inj} \lesssim 10^{42}$ erg s$^{-1}$ for \coolingflow) the X-ray luminosity from the galaxies is dominated by CGM itself, the contribution from the outflow becomes significant only if the energy injection rate is higher.}
    \label{fig:Lx-Edot}
\end{figure}

\subsection{Shock morphology in wind-driven vs jet-driven feedback}
In the AGN simulations, we have only explored energy injection methods that produce outflow in a wide angle (half opening angle $\theta_{\rm inj} = 45^\circ$) and is suitable for mimicking SNe and AGN wind feedback but may not be suitable for AGN jets that are highly collimated ($\theta_{\rm inj} \sim$ a few degrees). Even within the SNe/AGN wind driven cases, the morphology differs slightly based on the injection duration. For example, the morphology of the shock for the \nine case appears egg-like (considering the morphology in the other quadrants) as is expected, whereas, the shock in the \agninst case appears to be a dumbbell shaped (see Figure \ref{fig:SX-at-given-r})  with the center of each lobe offset from the galactic center.  Such behavior in the \agninst case is because the energy injection occurs only within $1$ Myr with a very high rate ($\approx 4\times 10^{44}$ \ergps; see Figure \ref{fig:inj_param}). The very high energy injection rate allows the shock to reach a distance of $\sim 8$ kpc within $1$ Myr for the assumed CGM parameters (see section \ref{sec:analytical}). Switching off of the energy source forces the AGN driven wind to deposit all its kinetic energy at that radius and, therefore, the subsequent evolution of the shock is roughly like a spherical blast wave but centered at $\sim 8$ kpc. This double-lobed behavior is consistent with the global galaxy simulations such as TNG-50 \citep{Pillepich2021}.

However, since the AGN jets are highly collimated, the produced shock would be quite elongated in the absence of any dissipation mechanism \citep{Begelman1989, Marti1997, Bromberg2011, Sarkar2022b}. Such shock expansion needs to be considered separately since the jet-ambient medium interaction is uncertain.  For example, if the collimated jets interact with dense clumps in the ISM, it may dissipate and thereby produce wind-like energy \citep{Mukherjee2018, RDutta2024}. In such a case, our current energy injection models would also be applicable to AGN-jet feedback.

\subsection{CGM luminosity versus feedback energy injection rate}
Observations of extra-planar X-ray emission from edge-on galaxies have been possible using \textit{CHANDRA}. \cite{Wang2016} obtained the $0.5-2.0$ keV X-ray luminosity ($L_X$) of extra-planar gas around 52 disk galaxies and found that the X-ray luminosity has a sub-linear dependence with the SFR. In Figure \ref{fig:Lx-Edot} we compare the total X-ray luminosity ($L_{X}$) of our model galaxy having different CGM models and feedback methods with the observations of \cite{Wang2016} for a galaxy of stellar mass $M_{\star} = 5 \times 10^{10}$ \msun. We calculated the X-ray luminosity, $L_X$, by integrating the surface brightness profile within the inner 30 kpc region. The red-shaded region shows the best-fit relation (with $1-\sigma$ scatter) between $L_{X}$ and SFR from \citet{Wang2016} for a Milky-Way-type galaxy, where we converted the SFR to energy injection rate using $\dot{E}_{\rm inj} = 4 \times 10^{41} \left(\frac{SFR}{{\rm M}_\odot \rm yr^{-1}}\right)$ erg s$^{-1}$ from \textsc{starburst99} as above. The circles, squares, and `plus' symbols represent the \isothermal, \isentropic, and \coolingflow\, CGM models respectively and the color represents different feedback methods. The horizontal grey dashed lines show the total X-ray luminosity of the initial CGM gas distribution in each case. The size of the symbols denotes the time, with larger sizes denoting later times. We see that the total X-ray luminosity calculated from our simulations shows a sub-linear relationship with the SFR, similar to the \textit{CHANDRA} observation.  At low energy injection rates of ($\dot{E}_{\rm inj} \lesssim 10^{41}$ erg s$^{-1}$ for \isothermal\, and \isentropic\, and $\dot{E}_{\rm inj} \lesssim 10^{42}$ erg s$^{-1}$ for \coolingflow) the X-ray luminosity is similar to the no-feedback solutions (horizontal dashed lines), indicating that the feedback contribution is subdominant, while at higher energy injection rate the CGM luminosity is above the no-feedback solution.

We note that this conclusion on when feedback dominates the X-ray emission from the CGM will not hold if the no-feedback CGM luminosity depends on the SFR. Such a dependence is expected at least in the cooling flow CGM scenario, where the X-ray luminosity scales with the inflow rate $\dot{M}_{\rm if}$, which in turn equals roughly half the SFR (the $\dot{M}_{\rm if}\approx0.5$~SFR relation is a result of the cooling flow being the source of fuel for star formation, see \citealt{Stern2020}). For the  $\dot{M}_{\rm if}$ = 1~\msun yr$^{-1}$ assumed in the \coolingflow\, initial conditions above, Fig.~\ref{fig:Lx-Edot} shows that $L_{X,{\rm soft}}=4\cdot10^{39}$ erg s$^{-1}$, and this luminosity is hardly increased after accounting for feedback from a corresponding SFR of 2~\msun yr$^{-1}$. This lack of change in X-ray luminosity is expected also for other values of the SFR since a different SFR will change both the no-feedback CGM luminosity and the feedback energy injection rate by the same factor. We thus expect that if star formation is steady and fueled by a cooling-flow in the CGM, the CGM X-ray luminosity will roughly equal that in the no-feedback solution.  

\subsection{Comparison with previous studies}
Several recent studies have focused on examining the interaction between outflows and the CGM, employing both idealized and cosmological numerical simulations. \citet{MLi2020}  and \citet{Vijayan2022} performed idealized simulations to investigate the effect of gravitationally bound hot outflows on CGM gas. \citet{MLi2020} injected the mass and energy corresponding to a feedback event at random positions three kpc above the galactic plane to mimic the random positions of star-clusters in the galactic disk, in contrast with our choice to inject at the galaxy center. They studied the effect of multiple star-forming events with constant SFR and with a 10 Myr time-gap between the SF events. They found that the outflows form a large-scale, metal-enriched atmosphere with the hot gas having both outward and inward velocity confirming a fountain-like motion. In this paper, we have focused on a single SF/AGN event and tried to understand how it changes the CGM gas properties. We noticed that even though the life span of a star-burst event is $\sim 40$ Myr, its effects last up to 200-300 Myr. So the shocks produced in subsequent bursts in the \citet{MLi2020} simulation traverse an already modified CGM and ultimately produce a fountain effect balancing the outflowing and inflowing gas in the galaxy. In another study, \citet{Ramesh2023} conducted an analysis of 132 galaxies resembling the Milky Way using the TNG-50 simulations. They noted that both increasing stellar mass and star formation rate (SFR) are associated with a prominent rise in X-ray luminosity. However, the correlation between SFR and X-ray luminosity is tighter for galaxies with a size smaller than Milky Way. In our idealized simulations, we find that X-ray luminosity increases with SFR but the correlation is sub-linear in nature. 

\section{Summary}
\label{sec:summary}
AGN black hole accretion and star formation activity in galaxies can potentialy give rise to large-scale outflows that drive a significant amount of gas and energy from the ISM into the CGM. In this paper, we carry out controlled simulations to study the effects of such outflows for Milky-Way-sized galaxies and investigate the X-ray response of the CGM to the outflows. We consider different hydrostatic CGM models for the initial CGM states; the \isothermal\, \citep{Faerman2017}, \isentropic\, \citep{Faerman2020}, and \coolingflow\, \citep{Stern2023} models and a range of energy injection rates  ($\dot{E}_{\rm inj} \sim 10^{40} - 10^{44}$ erg s$^{-1}$) for star-formation or black-hole accretion driven outflows to study the wind-CGM interaction. We perform hydrodynamical simulations using \textsc{pluto} \citep{Mignone2007} within a box size of 258 kpc in spherical coordinates with a logarithmically decreasing resolution of $ \sim 3$ pc at the inner boundary and $\sim 4$ kpc at the outer boundary. Although $\Delta r$ at the virial radius is notably higher than that typically found in state-of-the-art cosmological simulations, the relevant resolution for our simulations is $\Delta r\sim 0.4$ kpc (at $30$ kpc), comparable to the cosmological simulations. We also ran tests with $\Delta r \approx 15$ pc through out the simulation box and validated our results. Our main results are as follows:

\begin{itemize}

    \item When feedback energy rates exceed $\sim 10^{41}-10^{42}$ erg s$^{-1}$, with the thresholds increasing with the assumed initial gas density, outflows originating from the central region of the galaxies drive a shock wave through the CGM, forming a low-density bubble surrounded by a dense gas shell. The bubble size is influenced by CGM density, with higher densities leading to smaller bubbles. The strong feedback events modify the CGM gas distribution, with the hot dense region near the outer shock significantly enhancing X-ray emissions, creating a limb-brightened feature. Conversely, the region enclosed by the shell contributes less to X-ray emissions due to its gas scarcity. If feedback is strong enough, the projected X-ray profiles appear to be completely flat in the inner region due to the presence of the low density cavity.
    
    \item The observed surface brightness in galaxies (as observed by \erosita) is significantly higher than the \isentropic\, predictions. High surface brightness can be obtained in the \isentropic\, model if there is a strong central energy injection. However, \isentropic\, CGM model falls significantly short of explaining the high surface brightness at $\gtrsim 50$ kpc even with a strong energy injection (Figure \ref{fig:1d-xray-profiles}). On the other hand, the \isothermal\, and \coolingflow\, models are much closer to the observed surface brightness profiles and can produce X-ray bright CGM at $\gtrsim 50$ kpc with a reasonable energy injection at the center.
    \item The CGM density profile has a significant role in the propagation of the shock and the X-ray surface brightness. The initial \isentropic\, model has a comparatively lower density than the \isothermal, and \coolingflow\, models, hence the CGM is modified to larger distances for the initial \isentropic\, profile with the same amount of energy injection. The effect of the central energy injection is most visible in the \isothermal\, and \isentropic\, CGM profiles where the density is lower near the galaxy. Therefore, although the \isentropic\, CGM is initially below the \erosita detection level, a central energy injection can boost the X-ray brightness for the telescope to detect it. The \coolingflow\, CGM, on the other hand, is not affected much by the outflows unless there is a large amount of energy ($\sim 10^{58}$ erg) injected.
    
    \item For low energy injections, the X-ray surface brightness can not distinguish between the shocks created by the SNe or the AGN. For large energy injection cases, the SNe produce much more X-rays than its AGN counterparts. This is due to significant X-ray emission from the SNe ejected material and can be recognized by its double-peaked brightness in the radial X-ray profiles.
    
    \item The total X-ray luminosity ($L_X$) of our model galaxy, when subjected to a feedback event, reveals a sub-linear relationship with the star formation rate (SFR) or energy injection rate ($\dot{E}_{\rm inj}$), similar to the \textit{Chandra} observations of disk galaxies similar in mass to the Milky Way \citep{Wang2016}.
    
    \item We also present an analytic model, which describes the density of CGM as a power law and models the outgoing shock in the CGM using a model similar to a stellar wind. This model can predict the displacement of the outer shock in the CGM over time, and for the initial \isothermal\, and \isentropic\, model also approximates the diffuse emission obtained from the simulation. Additionally, our analytical model can be used to get constraints on the central energy injection and the CGM profile of a galaxy from the observed X-ray surface brightness.
    
\end{itemize}

Our computations represent initial steps in our understanding of how the CGM gas interacts with a single feedback event lasting a few million years. In more realistic scenarios, the CGM gas may undergo multiple feedback events over a timescale of $\sim$ few Gyr, while also accreting gas from the intergalactic medium. In future work we aim to investigate the long term behavior of the CGM in the presence of periodic feedback.

\section*{Acknowledgements}
We thank Aya Bamba, Yakov Faerman, Chris McKee, Yossi Oren, Stephanie Tonnesen, and Daniel Wang for insightful discussions and the referee for a helpful and constructive report. This work was supported by the German-Israel Foundation via Grant STE/1862-2 GE/625 17-1, by the Center for Computational Astrophysics (CCA) of the Flatiron Institute, and by the Mathematical and Physical Sciences (MPS) division of the Simons Foundation.

\section*{DATA AVAILABILITY}
The software and data employed in this study are referenced within the article.

\appendix
\section{X-ray emissivity}
\label{sec:emissivity_cloudy}
\begin{figure}
  \centering
  \includegraphics[width=0.5\textwidth]{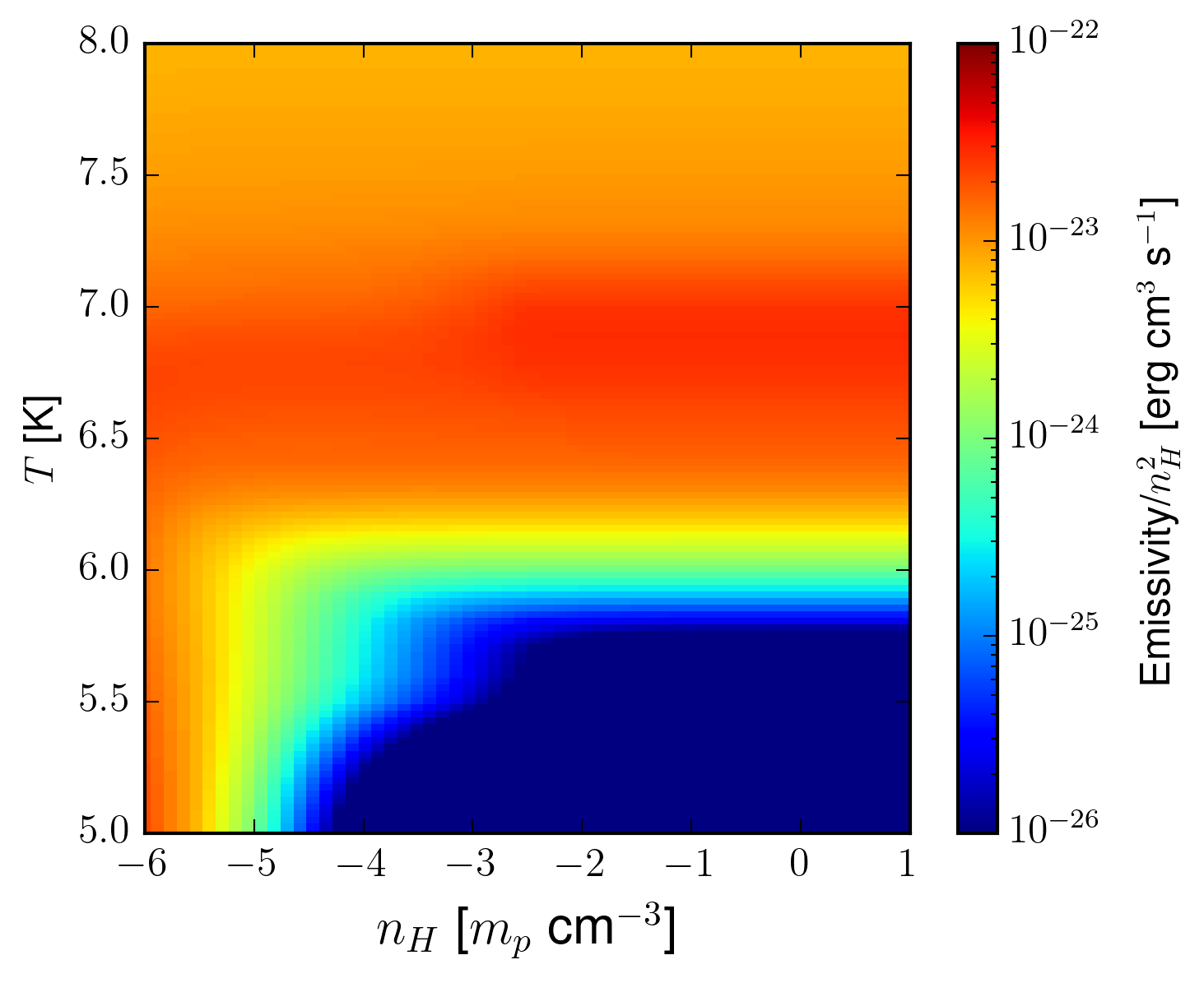}
  \caption{Soft X-ray ($0.5-2$ keV) emissivity for the plasma as a function of hydrogen density ($n_H$) and gas temperature, assuming solar metallicity. Calculations are done with \textsc{cloudy-v22}.}
  \label{fig:emis_solar}
\end{figure}

The soft (0.5-2 keV) X-ray emissivity ($\Lambda_X (n_H, T, Z)$)  has been calculated using \textsc{cloudy-v22} \citep{cloudy2023} assuming the UV background radiation of \cite{Haardt2012ApJ}. Figure \ref{fig:emis_solar} shows the X-ray emissivity (per $n_H^2$) for the Solar metallicity. For a purely collisionally ionized plasma, the emissivity goes to very low values at $T\lesssim 10^6$ K, as is the case for $n_H \gtrsim 10^{-4}$ m$_p$ cm$^{-3}$. However, for lower densities ($n_H \lesssim 10^{-4}$m$_p$ cm$^{-3}$), photo-ionization can result in a high enough abundance of highly ionized species that the X-ray emissivity becomes significant. Above $T\gtrsim 10^6$ K, the emissivity, however, is independent of the density and only a function of the temperature.

\section{Resolution test}
\label{sec:res_test}
\begin{figure}
  \centering
  \includegraphics[width=0.5\textwidth]{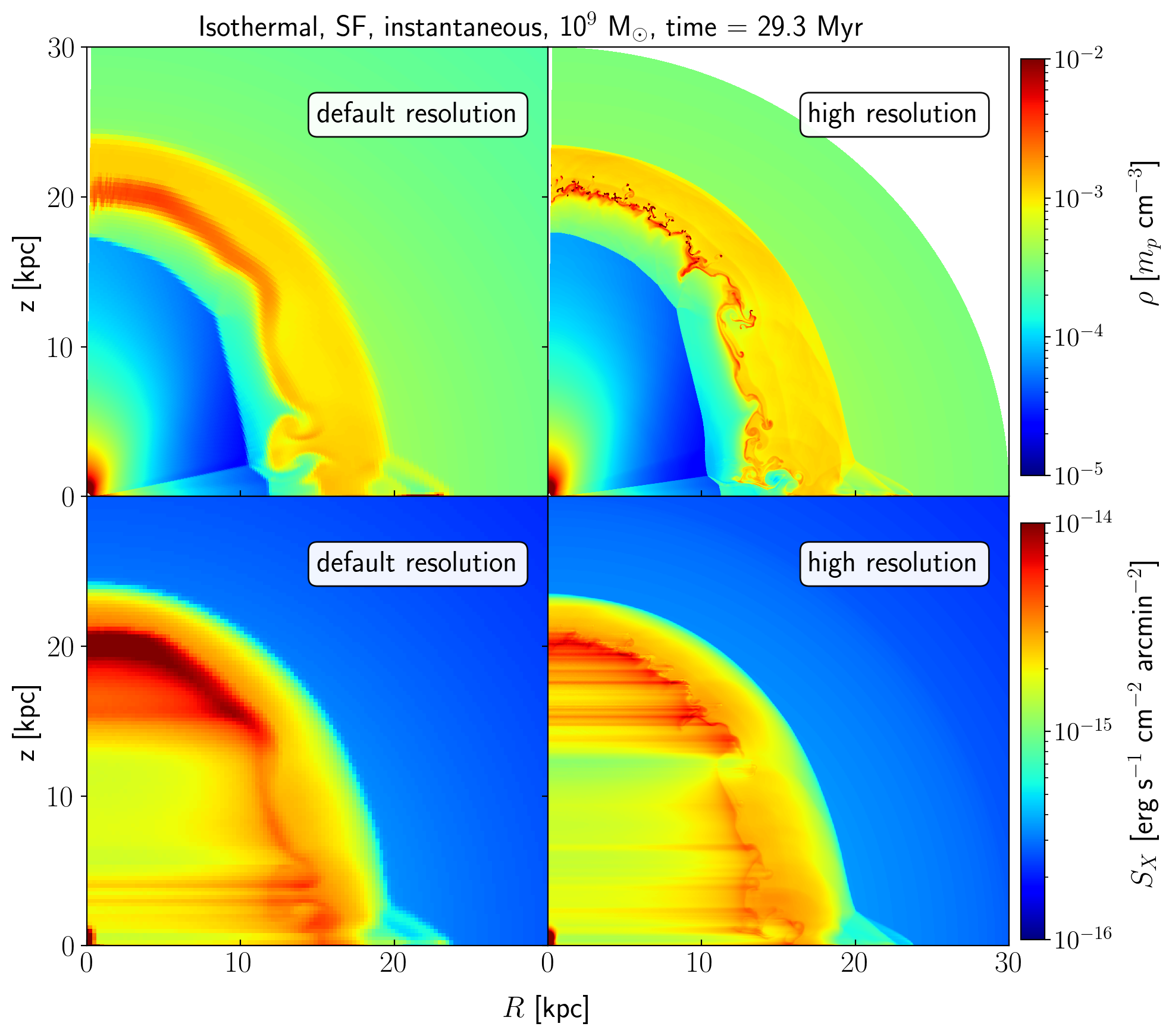}
  \includegraphics[width=0.5\textwidth]{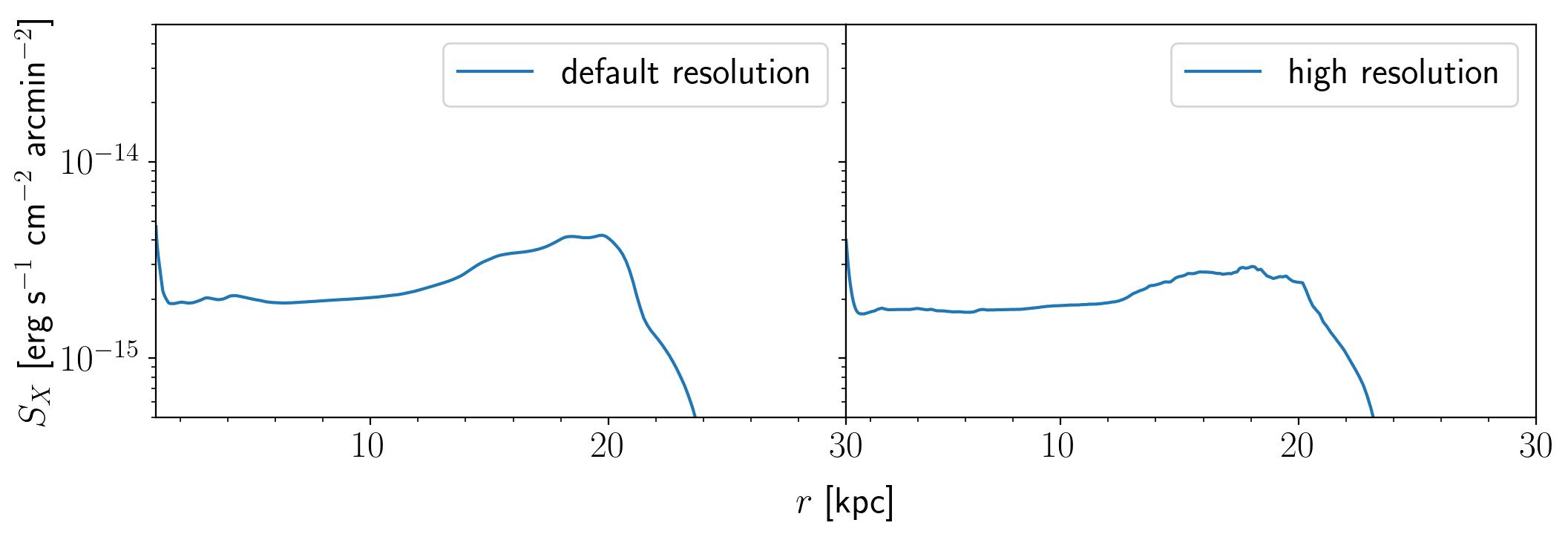}
  \caption{Resolution test for the X-ray surface brightness. The left panel shows a snapshot at $t=29$ Myr for the \nine model at the default resolution. The right panel shows the same simulation but with a grid resolution of $N_r\times N_\theta= 2048\times 1024$. We notice that the surface brightness at the peak changes by $\approx 40$\%.}
  \label{fig:res_check}
\end{figure}

We perform resolution tests for the \eight, \nine, \agninst and \agncont cases. The new simulation boxes extend only to $30$ kpc, unlike $258$ kpc in the fiducial simulations. We employ a uniform grid of $2048\times1024$ cells in the $r$ and $\theta$ directions, respectively. Although our high-resolution simulations do not include CGM beyond $30$ kpc, we consider its effect till $258$ kpc while calculating the surface brightness. A comparison of both these simulations for \nine is shown in Figure \ref{fig:res_check}. We see that a lot of the medium-density ($n\sim 4\times 10^{-3}$ \mpcc) gas in the contact discontinuity cools down and fragments into denser clouds. This fragmentation reduces the X-ray emissivity of the contact discontinuity. However, the reduction is by $\approx 40\%$ at the peak as can be seen in the figure. The high-resolution simulations for \eight and AGN cases do not show any significant differences compared to its fiducial resolution simulations (see Figure \ref{fig:SX-at-given-r}, bottom panel, for AGN cases).

Note that the horizontal line-like features in the 2D X-ray projection maps are not artifacts; they emerge because our simulations are 2D and axisymmetric. This means that any point in the 2D density and temperature distribution essentially represents a ring in the full 3D simulation. When we create a projected map of X-ray surface brightness, we consider a 3D sphere with a radius of $r_{\rm vir} = 258$ kpc. In the process of generating X-ray surface brightness maps from 3D simulations, we flatten the 3D data onto a 2D plane, causing the rings to transform into horizontal lines.

\section{Gravitational force for isothermal and isentropic CGM}
\label{sec:klypin}
The gravitational force for \isothermal\, and \isentropic\, CGM follows the \citet{Klypin2002}'s model and is given as 
\begin{equation}
    g(r) = -G\: \frac{M_b(r)+M_{\rm dm}(r)}{r^2}
\end{equation}
where $M_b(r)$ is the total baryonic mass and $M_{\rm dm}(r)$ is the total dark matter mass within radius r. \citet{Klypin2002} modeled each component of the baryonic mass, like nucleus, bulge/bar, disk, etc. separately but also provided an analytic profile of the baryonic mass in the galaxy
\begin{eqnarray}
\hspace{-3cm}
M_b(r) = \hspace{-0.5cm} && M_{\rm BH}\\ \nonumber
             && + 0.025\,M_b^{\rm vir}\,[1-\exp(-2.64 r^{1.15})]\\ \nonumber
             && + 0.142\,M_b^{\rm vir}\,[1-(1+r^{1.5}) \exp(-r^{1.5})) \\ \nonumber
             && + 0.833\,M_b^{\rm vir}\,[1-(1+r/r_d)\exp(-r/r_d)] 
\end{eqnarray}
where $r$ is the radius in kpc, $M_b^{\rm vir}$ is the total mass in cooled baryons and $r_d$ is the scale length of the disk.

The dark matter density profile is assumed to be an NFW mass distribution profile \citep{Navarro1997} 
\begin{eqnarray}
    \rho_{\rm dm}(r) &=& \frac{\rho_s}{x(1+x)^2}, \hspace{1cm}  x = r/r_s\\ \nonumber
    M_{\rm dm}(r) &=& 4\pi \rho_s r_s^3 f(x) \\ \nonumber
    &=& M_{\rm vir}f(x)/f(C) \\ \nonumber
    f(x) &=& \ln(1+x) - \frac{x}{1+x}
\end{eqnarray}
where $M_{\rm vir}$ and $r_{\rm vir}$ are the virial mass and radius of the galaxy. $C$ is the halo concentration parameter and $r_s = r_{\rm vir}/C$. The corresponding parameter values are given in Table \ref{tab:modelpara}.
\begin{center}
\begin{table}
\caption{Parameters used to describe the gravitational force}
\label{tab:modelpara}
\hspace{1.5cm}
\begin{tabular}{|c|c|}
\hline
\hline
Parameters (Units) & Values \\
\hline
$M_{\rm vir}$ (\msun) & $10^{12}$ \\
$M_b^{\rm vir}$ (\msun) & $6\times 10^{10}$ \\
$m_{\rm BH}$ (\msun) & $2.6 \times 10^6$\\
$r_{\rm vir}\,(\rm kpc)$ & $258$ \\
$r_{\rm s}\,(\rm kpc)$ & $21.5$\\
$r_d\,(\rm kpc)$ & 3\\
$r_s\,(\rm kpc)$ & 21.5\\
$C$ & 12\\
\hline
\hline
\end{tabular}
\end{table}
\end{center} 

\bibliographystyle{mnras}
\bibliography{reference}

\end{document}